\documentclass[fleqn,usenatbib]{mnras}

\usepackage{newtxtext,newtxmath}

\usepackage[T1]{fontenc}

\DeclareRobustCommand{\VAN}[3]{#2}
\let\VANthebibliography\thebibliography
\def\thebibliography{\DeclareRobustCommand{\VAN}[3]{##3}\VANthebibliography}

\usepackage{graphicx}	
\usepackage{amsmath}	
\usepackage{anyfontsize}

\title[Eruptive YSOs in Cygnus with NEOWISE \& SPICY]{Eruptive YSOs in Cygnus-X: a mid-infrared variability study with NEOWISE and SPICY}

\author[C. Morris et. al.]{C. Morris$^{1}$\thanks{E-mail: c. morris6@herts.ac.uk},
Z. Guo$^{2,3,1}$, P. W. Lucas$^{1}$,  N. Miller$^{4}$, C. Contreras Pe\~{n}a$^{5,6}$ and M. A. Kuhn$^{1}$
\\
$^{1}$Centre for Astrophysics Research, University of Hertfordshire, College Lane, Hatfield, Hertfordshire, AL10 9AB, UK\\
$^{2}$Instituto de F{\'i}sica y Astronom{\'i}a, Universidad de Valpara{\'i}so, ave. Gran Breta{\~n}a, 1111, Casilla 5030, Valpara{\'i}so, Chile\\
$^{3}$Millennium Institute of Astrophysics, Nuncio Monse{\~n}or Sotero Sanz 100, Of. 104, Providencia, Santiago, Chile\\
$^{4}$Department of Physics \& Astronomy, University of Wyoming, 1000 E. University, Laramie, WY 82070, USA \\
$^{5}$Department of Physics and Astronomy, Seoul National University, 1 Gwanak-ro, Gwanak-gu, Seoul 08826, Republic of Korea\\
$^{6}$Research Institute of Basic Sciences, Seoul National University, Seoul 08826, Republic of Korea\\
}

\date{Accepted XXX. Received YYY; in original form ZZZ}

\pubyear{2023}

\begin{document}
\label{firstpage}
\pagerange{\pageref{firstpage}--\pageref{lastpage}}
\maketitle

\begin{abstract}
The mass accretion process controls pre-main-sequence evolution, although its intrinsic instability has yet to be fully understood, especially towards the protostellar stage. 
In this work, we have undertaken a thorough examination of the mid-infrared variability of Spitzer-selected YSOs in the Cygnus-X star-forming region over the last decade, using the \textit{NEOWISE} time series. This work compares two groups of young stars: embedded Class I objects, and the more evolved flat-spectrum/Class II sources. We report on 48 candidate eruptive variables within these groups, including 14 with characteristics that resemble the photometric behaviour of FUors. We also include an additional 20 YSOs, which are of a less certain categorisation. We find the candidate FUors to be an order of magnitude more common among the younger Class I systems than more evolved objects. A large number of the identified short-duration eruptive YSOs display mid-infrared colour behaviour that is redder-when-brighter, which contrasts with optically bright outbursts seen in YSOs. Finally, we note the unusual long-term rising behaviours of four Class I YSOs, with rise timescales longer than five years, which is far slower than $\sim$6-12 month timescale for the majority of optically discovered FUors. Additionally, our broader investigation of MIR variability for embedded class I YSOs shows that there is a higher incidence of high amplitude variability for these stars, than is seen in class II sources. This holds true for all variable class I YSOs, not just 
the eruptive sources. 
\end{abstract}




\section{Introduction}

Pre-main-sequence evolution of low-mass stars consists of four stages (from Class 0 to III), defined by the relative brightness of the stellar photosphere and the circumstellar disc or envelope \citep[][]{1996GreenLada}. Class 0 and Class I sources are the earliest evolutionary stages, so-called protostars, during which the young star is embedded in its envelope and optically invisible. During the more evolved Class II stage, the young stellar object (YSO) is directly observable at optical wavelengths, accompanied by active accretion funnel flows linking the circumstellar disc and the stellar surface \citep{2007AA_Tau_Bouvier}. Additionally, the "flat-spectrum" stage (FS) is defined as the transitional stage between Class I and II, where the stellar photosphere and accretion disk share similar luminosity, namely a flat spectral energy distribution (SED). Quantitatively, these pre-main-sequence evolutionary stages were classified by the near to mid-infrared spectral index $\alpha$, as Class I ($\alpha > 0.3$), FS ($-0.3 < \alpha < 0.3$), and Class II ($\alpha < -0.3$). Previous photometric surveys have revealed that the mass accretion process on YSOs has an inherent instability \citep[e.g.][]{Stauffer2014, Venuti2019, Carlos2020}. However, most past works focused on the optical to near-infrared variability of Class II YSOs. Systematic monitoring of the accretion variability of protostellar cluster members will shed light on the accretion variability towards earlier evolutionary stages, where the average accretion rate is higher than in the FS and Class II stages \citep[see the models used in][]{2017Fischer_orion}. 

The long duration time coverage of mid-infrared (MIR) surveys, by Spitzer, WISE and the current NEOWISE \citep{2007spitzerCyg,2010Wise,2014Neowise}, provides the ability to find variable events on multi-year timescales. The nearby massive star-forming complex of Cygnus-X (at roughly 1.4 kpc, \citet{rygl2012parallaxes}) makes an excellent test site for studying the long-term behaviour of YSOs, because of the wide range of surveys it is included within. These surveys cover multiple wavelengths with a long time baseline, providing a high completeness of YSO candidates in the region. The large sample of \citet{Kryukova2014}, which identifies over 2000 candidate YSOs, utilised the full MIR coverage provided by \textit{Spitzer} to isolate YSO candidates on the basis of MIR colour; these sources are exclusively Class I YSOs with a 24 micron (\textit{[24]}) detection. The same sky region is also covered by NEOWISE, The UKIRT Galactic Plane Survey (UGPS) \citep{2008Lucas_UGPS}, the Two Micron All-Sky Survey (2MASS) \citep{2mass}, \textit{Gaia} \citep{2023GaiaDR3}, and the INT Photometric H$\alpha$ Survey (IPHAS) \citep{iphas,Drew2005}.

In addition, there is a large list of YSO candidates available in the Cygnus-X region, from the \textit{Spitzer}-based SPICY \citep{2021Kuhn_spicy} catalogue of YSOs. SPICY uses a mixture of traditional infrared colour cuts and random forest classifiers to produce a comprehensive list of Class II and flat-spectrum YSOs in the $613\deg^{2}$ covered by the existing cryogenic-era Glimpse surveys \citep{2003BenjaminGLIMPSE,2011Glimpse}. It contains a large number of Class I YSOs and a reduced number of Class III sources owing to the much lower IR excess of those stars. In this instance, there are 7281 additional YSOs, to those found in the aforementioned sample from the Cygnus-X Legacy Survey \citep{2010BeererCX, Kryukova2014}\footnote{The data release document can be found here: https://lweb.cfa.harvard.edu/cygnusX/}. Given the broad wavelength coverage of the region, there lies an opportunity to investigate the range and frequency of both high amplitude variability generally and eruptive variability specifically, for YSOs at different evolutionary stages. 

The post-reactivation NEOWISE survey provides an opportunity to study the long-term time-domain behaviour of YSOs. The NEOWISE survey covers the whole sky in both the 3.4 $\mu$m $W1$ and 4.6 $\mu$m $W2$ bands. At present, there have been over 10 years of continuous monitoring, at $\sim$6 month intervals, comprised of many single scans. Considering the higher brightness at 24 $\mu$m of the Class I protostars, our sample of Class I YSOs will preferentially target the younger objects. Such a search for eruptive variables (EVs) at the protostellar stage has not been previously carried out within Cygnus-X.     

Variability is a common characteristic among YSOs. This is often of a low amplitude nature (\textless 1 mag in optical), driven by small changes in extinction along line-of-sight, accretion rate, or by stellar rotation. Alternatively, unpredicted high-amplitude accretion events are observed on young stars, during which YSOs accrete a significant portion of their final mass \citep[see][]{1996_Hart_Ken_FUors}. 
These bursts have a myriad of different triggering mechanisms, with theoretical studies able to reproduce outbursts via: Gravitational Instability (GI) and Magneto-Rotational Instability (MRI) in the inner disk \citep{2016Kratt_Lod,2023Bourdarot}, thermal instabilities \citep{1994Bell_lin, Nayakshin2024a}, disk fragmentation, either due to GI \citep{2010Vorobyov}, or young massive planets \citep{2005Clarke}, stellar flybys \citep{2019Cuello,2022Borchert_Flyby} and evaporating young hot Jupiter \citep{Nayakshin2024b}. This wide range also ensures that there is no universal model for an episodic accretion event, and thus large observationally confirmed samples of these events will be required to understand them further.
Observationally, the young accretion bursts are traditionally characterised into two distinct groups, based on their duration and amplitude. The FUor-type, named after FU Ori, are long-lasting (order of decades or centuries) outbursts of high amplitude and a comparatively short brightening time scale, on the order of $\sim$1000 days \citep{1996_Hart_Ken_FUors}. These outbursts are considered rare, with only a few dozen confirmed by both spectroscopy and photometry \citep{Connelley2018}, including those recently identified by VVV and Gaia time series photometry \citep[such as][]{2017CarlosCP, Lucas2024, 2024Zhen_FU, Guo2024b, 2018Hillen_G17bpi}. More recent works \citep[such as][]{Carlos2023Wise, 2024Tran_10aaa, 2024Park} have made use of the NEOWISE time series to identify younger or more heavily embedded FUor candidates, with a view to increasing the number of systems known. The NEOWISE stacked image catalogue unTimely has also been used for this purpose \citep{2024LiWang}. Additionally the survey of \citet{Park_2021} quantified the array of variabilities seen in YSOs with NEOWISE (although the majority of those authors outbursting sources were class II YSOs). The recurrence timescales of FUors are on the order of $\mathrm{10}^{5}$ years \citep{2013Scholz}. These outbursts are driven by massively enhanced accretion (crushing the magnetic  field of the star), which will then cause the inner accretion disk to become self-luminous via viscous heating, and outshine the young star's photosphere, hence the distinctive spectrum of these objects \citep{2009ZhuA, 2011Hartmann, 2022Liu}. Unlike FUors, there is another type of young outbursting stars called EXors. These have a much shorter outburst duration of between months and a few years as well as a symmetric rising/cooling light curve morphology \citep{Lorenzetti2012}. Some EXors have repeated outbursting behaviour on a time scale of a decade \citep{2023Mike} and some others may be periodic \citep{2022Zhen_periodic}. Uniform amongst EXors is that the YSOs maintain magnetospheric accretion (i.e the YSO's magnetic field is not crushed by the accreting material, and thus boundary layer accretion continues), with spectra frequently containing H{\sc i} emission lines \citep{Carlos2017b, Zhen2020}. Recently, a much wider array of outbursting behaviours has been noted among eruptive YSOs \citep[reviewed by][]{2022_PP7_var}, consisting of YSOs with photometric features similar to FUors, but with spectroscopic traits more comparable to EXors \citep[see][]{2004BricenoMNor, 2007AcostaMNor}. The range of behaviours is further challenged by the near-infrared VVV survey \citep{Minniti2010, 2021_Zhen_MW}, where the majority of long-duration outbursts (\textgreater 10 years) were displaying EXor-like magnetospheric accretion signatures in their spectra. More recent literature has referred to intermediate duration outbursts as V1647 Ori-type \citep[see table 1 in][]{2022_PP7_var}, which do not fit into the classical EXor or FUor groups. V1647 Ori-type (also referred to as MNors) outbursts are of interest in this work, owing to their similarities to both FUors and EXors. These similarities pertain to both their intermediate duration (faster than those of classical FUors but longer than EXors) and their mixed spectral features (often featuring emission lines). The durations are of further note because of the number of newly discovered FUors with \textless 10 year outbursts. See \citet{2004BricenoMNor,2009AspinV1647,2013NinanV1647, 2021Park_V899,2022Park_EXlup} for further discussion on the group and of V1647 Ori in general.

The incidence rate of eruptive variability in YSOs is a subject of importance when considering the 'protostellar luminosity spread problem', but the variance in amplitude (with respect to the photometric bandpasses being used for observations) has not been probed in detail previously. Statistical studies using large-scale photometric surveys revealed that accretion outbursts are more common at younger stages, with a lower occurrence rate towards more evolved systems \citep[see][]{Contreras2024}. However, driven by the larger amplitudes observed in the optical and near-IR (as compared to those found in the MIR), YSOs with a less massive envelope are easier to detect during the outburst, leading to a bias on the estimated number density. 

Most eruptive YSOs have amplitudes $\geq$2 mag in NIR surveys \citep[see][]{2024Zhen_FU}, although there is a significantly higher ceiling than this (e.g. $\Delta W2 = 8$ mag \citet{2020Lucas_WIT}). 
We calculated amplitudes of a large number of confirmed eruptive variables (EVs) from the Outbursting YSOs CATalogue (OYCAT; Contreras Pe\~na et al., in prep).
We find that eruptive variables in the optical/NIR bandpasses have a median amplitude of 3.3 mag, 4.0 mag, 3.2 mag in the $G$, $R$ and $K_s$ filters, respectively.

A sample of nearby YSOs, selected from the MIR NEOWISE photometry, is desired to examine the range of eruptive behaviours towards the younger evolutionary stage. We have performed a detailed survey to monitor the long-term time-domain behaviour of these young systems, providing insight into how common the traditional FUor and EXor variables might be among a MIR-selected population, or if other eruptive events will dominate.

In this work, we will present two selections of YSOs within the region of Cygnus-X. Sample (1) consists of embedded objects with a 24 $\mu$m detection, mostly Class I YSOs with a small number of FS stars. Sample (2)  is comprised of non-embedded flat-spectrum and Class II/III sources (likely without envelopes), taken from The Spitzer/IRAC Candidate YSO Catalogue\citep[SPICY;][]{2021Kuhn_spicy}. The details of our sample selection methods and data processing procedures are presented in \S\ref{sec:methods}. In \S\ref{sec:results} we will discuss individual eruptive variables of interest within each selection. Then in \S\ref{sec:discussions} we will compare the incidence of eruptive variability (and of the FUor phenomenon in particular) between both populations and of similar sources from the literature. 

\section{Sample Selection and Data Processing}\label{sec:methods}

In this section, we will present the data products applied in this work, including archived data products from the WISE satellite and PI observations for highlighted sources. We will also introduce our selection methods to identify eruptive YSOs among the members of the Cygnus-X star-forming region. 

\subsection{Archived WISE data}\label{ssec:WISE_LCs}

We obtained NEOWISE light curves from the NASA/IPAC InfraRed Science Archive for the 2007 YSO candidates in Cygnus-X \citep{2014Kryu}, among which 1552 sources have a detection in at least one epoch of the time series, in both \textit{W1} and \textit{W2}. Of these 1552 sources, 1332 (these forming our first sample) have detections in at least 50\% of the possible epochs of single scan data. We use the terms `scan' and `epoch' with regard to the NEOWISE time series to refer to the multiple single images (scans) that are taken over the course of a few weeks, this being repeated $\sim$2 times per year (epochs). For each of the 1332 light curves, the NEOWISE single-scan data were combined with the earlier WISE measurements (ALLWISE, WISE 3-Band Cryo, and Post-Cryo, where available, acquired from the ALLWISE Multi-Epoch Photemetry Table). The photometry from the individual scans was then cleaned by removing detections with a high psf error. Additionally, those points with either \textit{w1snr} or \textit{w2snr} \textless 5 were subject to cuts first used in \citet{Koenig2014}:
\begin{equation}
    w1rchi2 < \frac{w1snr - 3}{7},
\end{equation}
\begin{equation}
    w2rchi2 < (0.1\times w2snr) - 0.3.
\end{equation}

Data points that did not pass either criterion were not included in the per epoch median magnitudes.
The error bars were taken as the standard deviation for the cleaned single-scan images at each epoch, rather than the stated photometric error in the catalogue. This change was motivated by the larger spread in fluxes at each epoch than the stated errors, which may be indicative of an underlying underestimate of errors in the NEOWISE data. In this work, we measured the standard deviation of the aforementioned spread to be 0.22 mag for the $W1$ bandpass, and 0.15 mag in $W2$. These values are obtained by first finding the average scatter in each `epoch' for a given star, then retrieving the standard deviation in this value across the whole sample. There is no clear magnitude dependence on the scattering within our sample. We also note that since most of our targets are faint sources located in crowded fields, they have systematically larger spreads per epoch compared with the brighter and more isolated WISE sources \citep[see][as an example]{Lucas2024}. Additional data cleaning involved removing photometric measurements associated with quality control flags attributed to the frame itself ($\mathrm{'qual\_frame'}\neq0$) or those had measured fluxes as upper limits only.  

To remove any bias associated with the selection of more heavily embedded YSOs (as well as investigating the distribution of EVs within this regime), a counterpart sample of sources was selected from the SPICY catalogue \citep{2021Kuhn_spicy}, featuring all candidate YSOs within the sky area used by \citet{2014Kryu}, that did not feature in those authors' sample. This SPICY sample includes 7281 stars that were not recognised by \citet{2014Kryu}, of which 5592 possess NEOWISE light curves, and 4935 have data of sufficient quality for analysis. We deem that a star must have data in at least 40\% of epochs to reliably analyse in this regard.  

Both the binned (the median of each scan within a single observation window) and unbinned light curves in the first sample (1332) were visually inspected to locate eruptive YSOs. Eruptions were considered for stars with amplitudes of over 1 magnitude, in either WISE bandpass (corresponding to $\sim$2 mag in optical band-passes), and obvious brightening from a quiescent state, leaving 299 light curves. Some sample members were further examined via the inspection of their unTimely \citep{2023unTimely} light curves, generated with the untimely catalogue tool \citep{2022Kiwy_UTCT}. The untimely catalogue is the time series of unWISE \citep{2018Meisner_unWISE}, which stack the individual NEOWISE sky-passes to create deeper ($\sim$ 1.3 mag) per epoch images, using the \texttt{crowdsource} source detection and photometry software \citep{2018Schlafly_DECam_CS}. The sources examined here were those with clearly variable light curves but with either high photometric error or a number of non-detections.
 
Candidate long-duration eruptive YSO outbursts were identified by fitting the light curves to a characteristic shape, defined by the rising slope, and a post-outburst linear decay. The two-component rising slope is defined in equations \ref{eq:FUorSlope1} \& \ref{eq:FUorSlope2}, as listed in \citet{Lucas2024}: 
\begin{equation}\label{eq:FUorSlope1}
    m(t)=m_{q}-\frac{s}{1+e^{-(t-t_{0})/\tau}}~~\mathrm{where}~~t~<~t_{0},
\end{equation}
\begin{equation}\label{eq:FUorSlope2}
    m(t)=m_{q}-s(0.5+0.25(t-t_{0})/\tau)~~\mathrm{where}~~t_{0}~\leq~t\leq t_{0} + 2\tau,
\end{equation}
where m(t) is magnitude as a function of time ($t$), $m_{q}$ is the magnitude at quiescence, $t_{0}$ is the time at which the star is at half the peak amplitude ($s$), and $\tau$ is the e-folding timescale. The fitting combines the above formalism with a linear decay that runs from $t_{0}+2\tau$ to the end of the time axis. The fit itself makes use of \texttt{scipy}'s \texttt{curvefit} routine, using the above model, to fit the initial guesses for five free parameters ($\tau$, $t_{0}$, $m_{q}$, $s$ \& the final magnitude), that are then fit by \texttt{emcee} over 10000 iterations (this code is now packaged as part of \texttt{aptare}\footnote{Available at https://github.com/nialljmiller/Aptare}). The sources with the lowest reduced mean squared error are then inspected to confirm as the FUor-type (examples are presented in Appendix \ref{app:fits}). We employed a cut-off value in root mean-squared error (RMSE \textgreater 0.1) between the fit and the normalised light curve to separate the candidates. This value is calculated from flux normalised between 0 and 1, and not the original apparent magnitude, and the RMSE is a measure of the average deviation (in this instance not more than 10\% of the amplitude). We note that this cut did lead to 2 likely genuine eruptive targets (considered as such from their amplitudes and overall morphology) failing the cut, which were re-confirmed by visual inspections. This method does struggle to identify sources with bursts at the start of the NEOWISE monitoring window, as well as those that are currently rising (or very recently finished), such as Source 2003. It should be noted that all stars that were identified as EV candidates by visual inspection were still subject to analysis and that this fitting method was trialled to locate YSOs resembling classical FUors. 

We also utilised the two-epoch UGPS survey \citep{2008Lucas_UGPS, Lucas2017} to identify YSOs with earlier outburst events (these being between 2011 and 2012) than would be seen between the epochs of the \textit{Spitzer} Cygnus-X Legacy Survey and ALLWISE/NEOWISE surveys. We selected members of the \citet{2014Kryu} catalogue featuring a detection in 50\% of UGPS images, then visually inspected the images to confirm genuine variability. This search identified 15 candidates, of which 11 were of sufficiently high amplitude ($>2$~mag in $K$) to be considered eruptive. Of those 11, only two were not picked up by the NEOWISE search method (see \S\ref{sec:results}). 

The techniques detailed in this section were focused on identifying FUor-type behaviour primarily, because of their rarity and the long timescale of these outbursts. EXor-type YSOs were still important in this work and will be discussed further below. The fitting methods above do not suit the shorter outbursts of these stars and would be challenging to distinguish from quasi-periodic and aperiodic dippers in many (but not all) cases. 

\subsection{Additional Data Products}

We present three $K$-bandpass spectra of two candidate EVs identified via the large flux differences between two photometric epochs of the UKIDSS survey \citep{Lucas2017}. 
Two spectra were obtained by the Near-Infrared Integral Field Spectrometer (NIFS) mounted on the Gemini North telescope \citep{McGregor2003}, in June and December 2013. Another spectrum was observed in July 2017 with the Infrared Camera and Spectrograph (IRCS) installed at the Subaru telescope
\citep{Tokunaga1998, Kobayashi2000}. The IRCS spectrum was reduced using the standard \texttt{twodspec} routines within \texttt{iraf}, and it has a resolution of 6.1\r{A} per pixel. The NIFS spectra were originally reduced by and part of the published PhD thesis of Carlos Contreras Pe\~{n}a \citep{2015PhDCarlos}.

\begin{table*}
	\centering
	\label{tab:Embedded_EVs}
 \caption{List of outbursting Cygnus-X members from \citet{2014Kryu}.}
	\begin{tabular}{lcccccccc} 
 \hline
		\hline
		Source & RA (deg) & Dec (deg) & $\Delta W1$ & $(W1 - W2)_{\rm min}$ & Spectral Index & Timescale & Outburst Duration  & Outburst Rise Time \\
  		Number & J2000  & J2000 & mag  & mag & $\alpha$ & Class & yr & yr \\
		\hline
            12$^{*}$  & 304.3677 & +41.9736 &  2.004 $\pm$ 0.019  & 1.01 & 1.03 & SDE & - & -\\ 
            121$^{*,\ddagger}$  & 305.5600 & +37.4743 &  1.848 $\pm$ 0.038  & 1.73 & 1.61 & SDE & - & -\\ 
            185  & 305.8536 & +37.5706 &  2.037 $\pm$ 0.07  & 3.05 & -0.08 & SDE & - & -\\ 
            232  & 306.0890 & +42.2683 &  1.692 $\pm$ 0.01  & 1.54 & 0.44  & SDE & - & -\\
            257$^{\dagger}$  & 306.1356 & +37.8303 &  3.279 $\pm$ 0.013 & 1.65 & 1.23 & LDE & 24 & 2\\   
            294$^{*}$  & 306.2141 & +38.4203 & 2.524 $\pm$ 0.020 & 0.77 & 1.2 & LDE & $>$30 & 7\\
            333$^{\ddagger}$  & 306.3550 & +39.3400 & 2.055 $\pm$ 0.011  & 1.33 & 0.4 & SDE & - & -\\ 
            352$^{*}$  & 306.4325 & +38.1869 & 1.795 $\pm$ 0.020 & 2.82 & 1.52 & SDE & - & -\\ 
            362$^{**}$ & 306.4640 & +39.3760 & 2.402 $\pm$ 0.023 & 1.86 & 0.58 & LDE & $>$10 & ? \\
            387$^{*}$  & 306.5333 & +41.6357 &  2.437 $\pm$ 0.097  & 1.97 & 0.27 & SDE & - & -\\
            397  & 306.5526 & +39.2783 & 2.262 $\pm$ 0.020 & 2.414 & 1.14 & IDE & 7-10 & $>3$\\
            441  & 306.6371 & +37.7786 & 1.674 $\pm$ 0.010  & 1.55 & 0.44 & SDE & - & -\\ 
            456  & 306.6672 & +40.0418 & 0.978 $\pm$ 0.016  & 0.7 & 0.16 & SDE & - & -\\
            591  & 306.9916 & +40.1975 & 2.354 $\pm$ 0.069  &  1.664  & 0.72 & LDE & $>$30 & 2\\
            625  & 307.0390 & +40.1518 & 3.371 $\pm$ 0.031  & 1.32 & 0.2 & IDE & 4 & $<$1\\
            658$^{\ddagger}$  & 307.1304 & +41.8473 & 1.749 $\pm$ 0.011 & 1.26 & 0.27 & IDE & 5 & $\approx$1\\ 
            769  & 307.4425 & +39.2864 &  2.817 $\pm$ 0.050  & 1.95 & 0.66 & SDE & - & -\\
            812  & 307.5106 & +40.3149 &  3.571 $\pm$ 0.046 & 1.93 & 1.54 & IDE & 6 & $<$0.6\\
            880  & 307.7106 & +41.2442 &  2.634 $\pm$ 0.030  &  1.27  & 0.22 & LDE & $>$10 & ?\\
            912  & 307.7863 & +40.0635 &  2.860 $\pm$ 0.036  &  2.732  & 0.88 & LDE & $>$30 & 5\\
            1017  & 308.0107 & +40.3108 &  2.313 $\pm$ 0.165 & 2.39 & 1.14 & IDE & 6 & $\approx$1.5\\ 
            1048  & 308.0877 & +41.1318 &  3.981 $\pm$ 0.511 &  1.733  & 1.7 & LDE & $>$30 & 7\\
            1475  & 309.0957 & +39.6769 &  2.210 $\pm$ 0.053 & 2.81 & 1.19 & LDE & $>$30 & 1\\ 
            1562$^{\ddagger}$  & 309.3483 & +42.2458 &  3.081 $\pm$ 0.062 & 2.49 & 1.01 & IDE & 3 & $<$1\\ 
            1626  & 309.5138 & +42.5483 &  3.081 $\pm$ 0.008 & 1.16 & 0.09 & SDE & - & -\\
            1738  & 309.8197 & +42.2693 &  2.401 $\pm$ 0.027 & 2.49 & -0.05 & Amb & ? & ?\\
            1769$^{*,\ddagger}$  & 309.9753 & +42.0143 &  2.878 $\pm$ 0.008  &  2.27 & 1.15 & Amb & ? & ?\\ 
            1884  & 310.3783 & +41.9181 &  1.438 $\pm$ 0.01  &  0.569  & 1.77 & IDE & 5.5 & 3\\ 
            1945  & 310.6990 & +42.7017 &  3.26 $\pm$ 0.041  &  1.961  & 0.77 & LDE & 10 & 3\\
            1964$^{**}$ & 310.8678 & +42.8167 & 0.767 $\pm$ 0.007 & 1.517 & 0.82 & LDE & $>$10 & ? \\
            1991$^{*}$  & 310.9918 & +42.8032 &  3.035 $\pm$ 0.056  &  2.827  & 0.96 & LDE & 9? & 7\\ 
            1999  & 311.0531 & +41.6165 &  4.606 $\pm$ 0.158 & 2.35 & 0.96 & IDE & 8 & 1.8\\ 
            2003  & 311.0754 & +41.6140 &  5.089 $\pm$ 0.017 & 1.67 & 1.01 & Amb & ? & ?\\ 
		\hline
  \hline
	\end{tabular}
 	
  \flushleft{LDE: long-duration eruptive sources with fast-rising FUor-type morphology and/or with outburst duration $>10$ yr.\\
    IDE: intermediate-duration events with a duration between 2 - 10 yrs.
    SDE: short-duration events including stochastic and quasi-periodic eruptive sources.\\
    Amb: ongoing outburst with ambitious duration. $^{\dagger}$: Source 257 is included in {\citet{Carlos2023Wise}}.\\
   $^{*}$: the \textit{Spitzer} detection is a blend between two detectable sources in UGPS. $^{**}$: Sources previously found by the two-epoch UPGS photometry.$^{\ddagger}$: Source has a light curve morphology that is hard to be certain of an accretion driven outburst.
   }
\end{table*}

\section{Results}\label{sec:results}

We present 68 candidate eruptive variables (EVs), across both the groups, those with MIPS $[24]$ detections and those SPICY selected stars without them. They are to be split into three classifications based on their outburst timescales (which are predicted in instances where the latest event is ongoing). The three groups are: `Long-Duration' of \textgreater 10 years (LDE), `Intermediate-Duration' of $2\leq t \leq 10$ years (IDE), and `Short-Duration' which are \textless 2 years (SDE). Overall we identified 13 `Long-Duration' YSO outbursts, with 9 of `Intermediate-Duration' and 41 `Short-Duration' events  (these events were solely identified by eye). A final five of the YSOs with ongoing outbursts at the final epoch had not been active long enough to classify in this way, and are listed as `ambiguous'. These timescale-based classifications are based on those used in \citet{Contreras2024}, although the cuts for short and intermediate durations have been slightly modified (less and greater than one year has now been changed to two years). This change was carried out to reflect the slightly variable sampling rate of NEOWISE making it somewhat challenging to reliably determine if an outburst is less than a year or a year and a half.

\subsection{Stars with MIPS [24] detection}\label{ssec:embedded_sources}

There are 33 EVs that have MIPS $[24]$ detection (which can be viewed in Table \ref{tab:Embedded_EVs}), and have an ID derived from the Vizier row number for the \citet{2014Kryu} catalogue. Most of the stars are Class I YSOs (28) and a small number of FS sources (5). The FS sources in this sample require $[3.6] - [4.5] \geq 0$, as per \citet{2014Kryu}. These sources were sorted into the three aforementioned categories based on the outburst timescale, as 11 long-duration, 8 intermediate-duration, and 11 short-duration outbursts. There are a final 3 stars that are of an ambiguous categorisation: sources 2003, 1738, \& 1769. The former has a notably high amplitude ($>5$~mag in $W1$), but as per the latest NEOWISE release the outburst is still ongoing, and thus of an unknown type and duration. By contrast, Sources 1738 \& 1769 have had multiple events observed in their light curves. For both systems, one burst appears as a long/intermediate duration type (ongoing as of 2024), and the other was a short-duration burst. Two additional stars (Sources 362 and 1964) are included in this group which were detected as candidate eruptive stars via the two-epoch UGPS survey \citep{Lucas2017}. All eruptive targets can be found in Table \ref{tab:Embedded_EVs}. The timescales in question are estimates, with two methods used: for SDEs and IDEs these are taken from the time between the start of the event and the point of return to the quiescent magnitude. For the LDEs the timescale is the fitted $\tau$ from equations \ref{eq:FUorSlope1} and \ref{eq:FUorSlope1}, unless the star has a plateau-like post-rise feature, in which case the timescale is stipulated to be greater than 10 years long.

\begin{figure}
    \includegraphics[width=0.95\columnwidth]{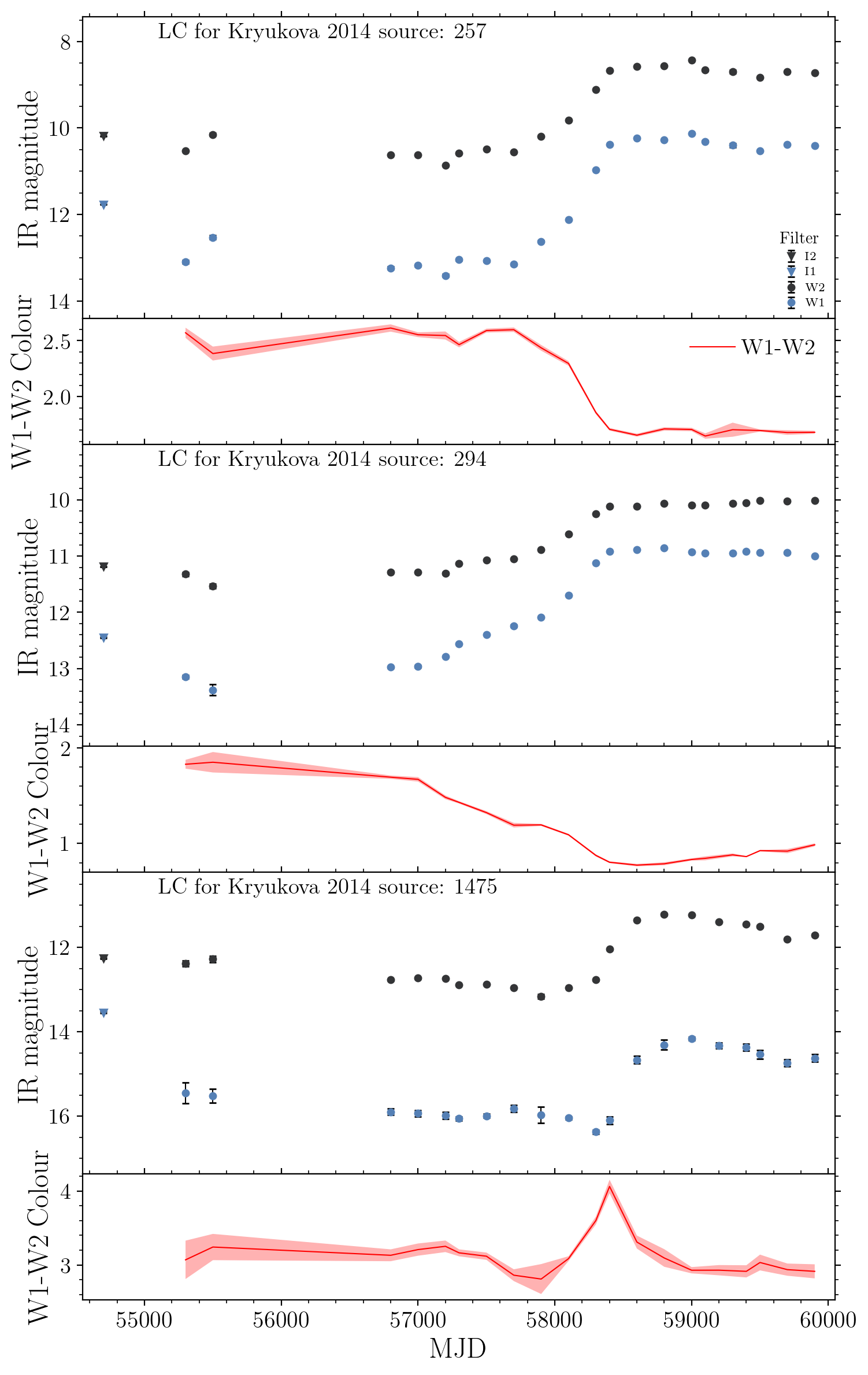}	
    \caption{NEOWISE $W1$ \& $W2$ light curves for three of the candidate eruptive variables with FUor type light curve morphologies. Each plot also contains a $W1-W2$ colour indicator (red line), with the error included as the pink-shaded region.}
    \label{fig:FUors_kryu_OB}
\end{figure}
\subsubsection{Long-Duration Eruptive YSOs}\label{sssec:E_FU}

Within the group of 11 eruptive variables with variability timescales greater than ten years, we note the presence of four stars with morphologies with similarities to those of classical FUors (Numbers: 257, 294, 591, \& 1475, see Figures \ref{fig:FUors_kryu_OB} \& \ref{fig:Kryu_LTEs_App}) i.e with a high amplitude outburst, that increases in brightness comparatively fast when compared to the following long duration decay. All display a high amplitude increase in luminosity over a 1 - 5 yr period, followed by a plateauing that appears set to continue long into the future. Sources 257 and 294 show no signs of post-outburst fading, whereas source 1475 has an estimated fading timescale of $\approx$ 24 years assuming a linear decaying slope.

Source 257 appears of a V1647 Ori-type on the strength that it has the characteristic emission spectrum in contrast to a FUor-type observed outburst. This source has been published in \citet{Carlos2023Wise}, under the identifier SPICY 109331. This star makes an excellent test case for the other members of this group, with such features as a high-amplitude (3.28 mag) burst in $W1$, rising over a short time-frame of $\sim2$ years, and becoming steadily bluer in its MIR colour, which is also seen in both FUors and V1647 Ori objects. 

\begin{figure}
    \includegraphics[width=0.95\columnwidth]{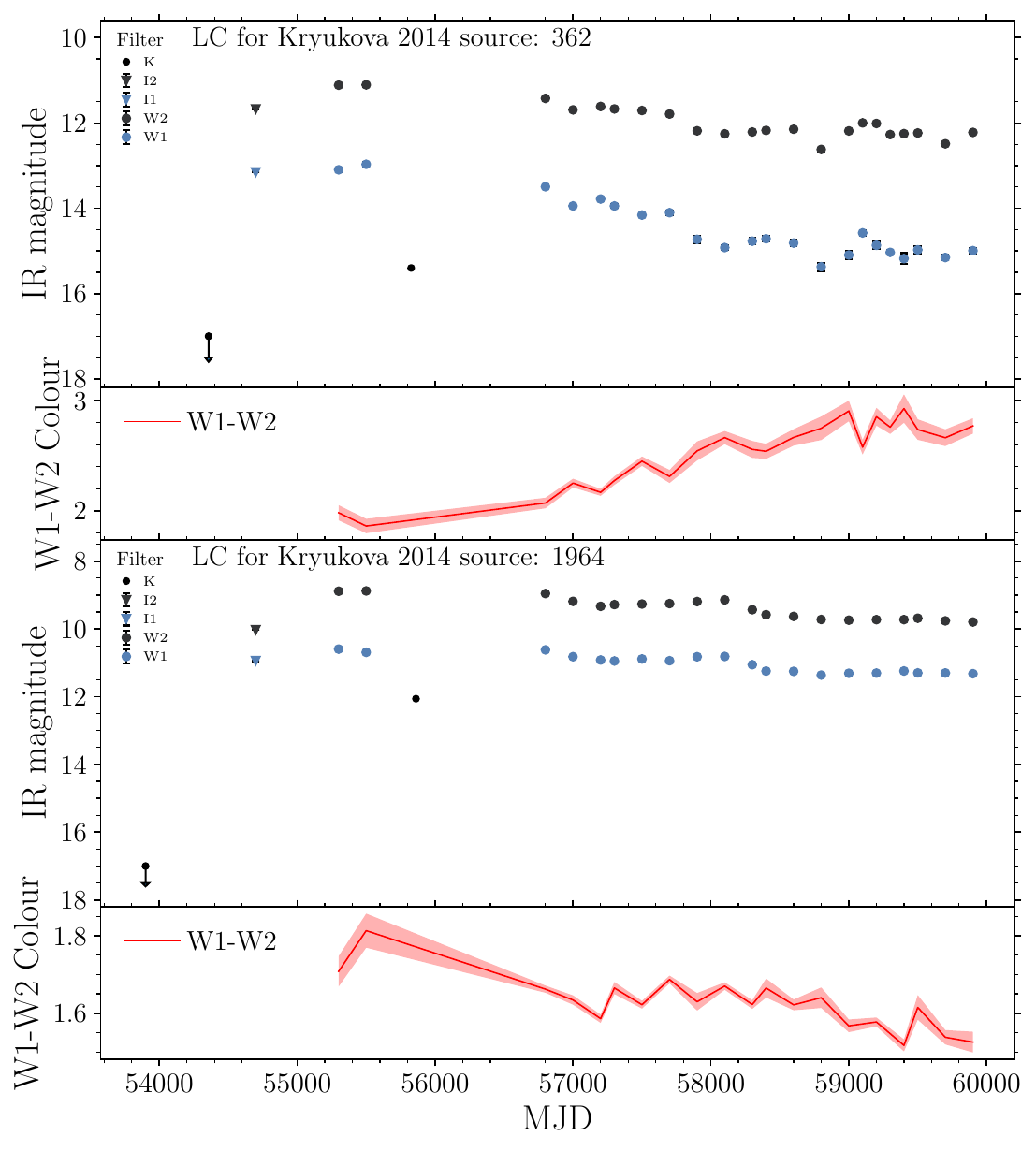}
    \caption{NEOWISE $W1$ \& $W2$ light curves for the two LDE candidates identified with UGPS photometry. The arrows represent the upper limits of UGPS for each field.}
    \label{fig:FUors_kryu_POB}
\end{figure}

In addition to the sources identified by their MIR eruptive behaviour, two further LDE candidates were identified on the basis of outbursts observed between the UKIDSS UGPS \citep{Lucas2017} NIR images (sources 362 and 1964). Both were identified by searching for bursts of at least 2 mag, wherein one UGPS epoch had a non-detection and the other had a minimum brightness of 15 mag (2 mag above the lower limit of the survey). In each case, the lower limit for the non-detection was calculated for the relevant UGPS image, using the \texttt{ImageDepth} tool within the \texttt{Photutils} package\footnote{By Larry Bradley et al. (2021), see \url{https://doi.org/10.5281/zenodo.5525286}} (lower limits being 16.9 mag, and 17.0 mag for sources 362 and 1964 respectively). The post-outburst light curves are displayed in Figure \ref{fig:FUors_kryu_POB}, both of which display the traditional fading behaviour of post-outburst FUors.

\begin{figure}
    \includegraphics[width=0.95\columnwidth]{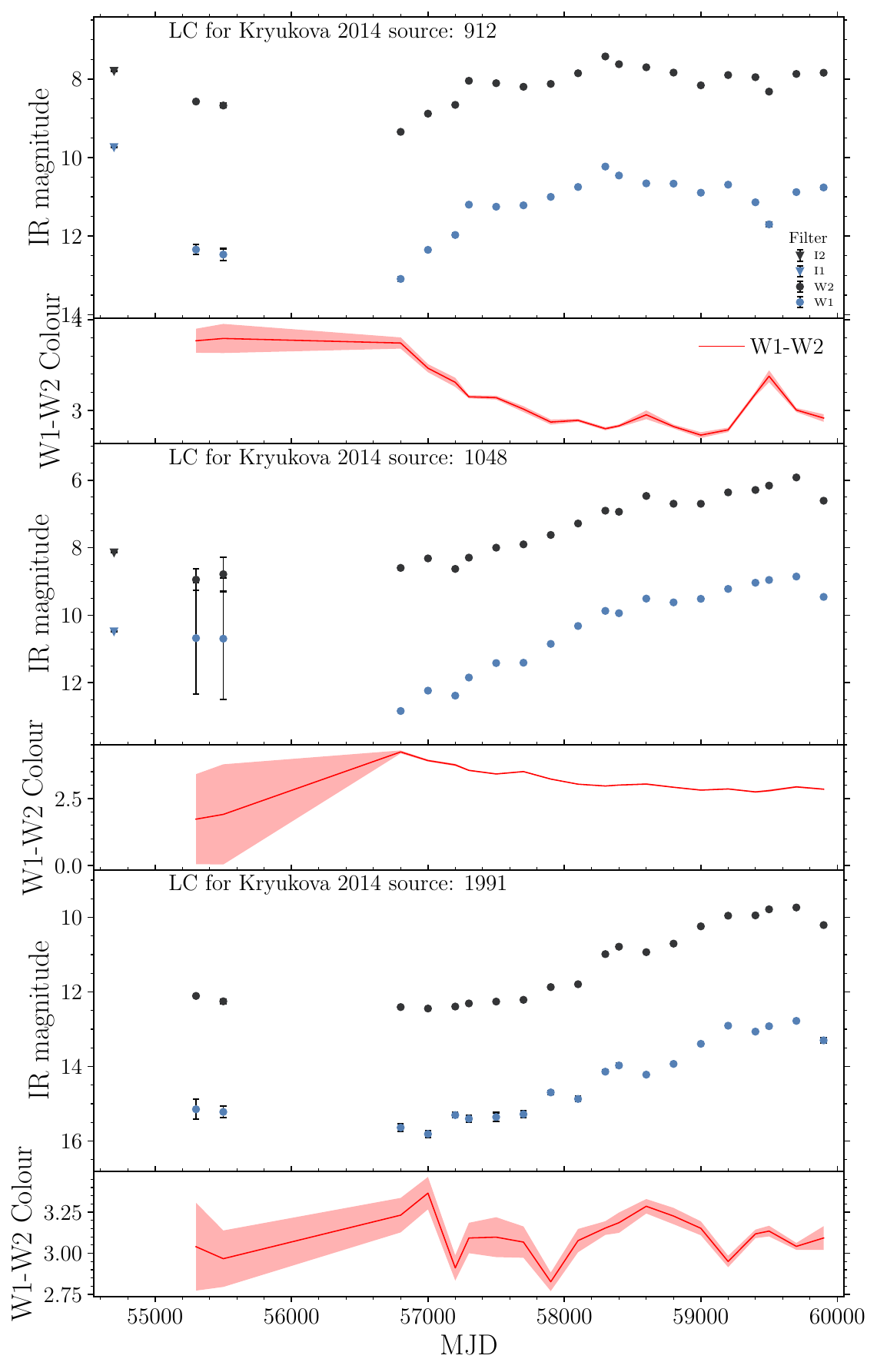}	
    \caption{NEOWISE $W1$ \& $W2$ light curves for three of the candidate eruptive variables with longer duration rise times. Each plot also contains a $W1-W2$ colour indicator (red line), with the error included as the pink-shaded region.}
    \label{fig:FUors_kryu_LT}
\end{figure}

A second group within the LDEs are stars with extended rising times (greater than 3 years) before reaching their photometric maxima, these are Sources 912, 1048, and 1991 in Figure \ref{fig:FUors_kryu_LT} (as well as Source 880 in Figure \ref{fig:Kryu_LTEs_App}). It is worth noting that sources 880 and 912 do seem to have faded from earlier observations, so there is a chance that these two systems may be 'dippers'. These slow-rising infrared outbursts have been observed infrequently before (albeit prior to a higher amplitude outburst), such as the FUor Gaia17bpi \citep{2018Hillen_G17bpi}, SSTgbs J21470601+4739394 \citep{2024Ashraf_SST}, and VVVv721 \citep{2017CarlosA,2017CarlosB}. This may suggest that the trigger for the outburst is driven by MRI (see \citet{2023Cleaver} for simulations of the low-power Gaia17bpi-like FUor outbursts).  Outbursts with rising times on the order of a decade (as seen here) are predicted in the flyby-triggered models of \citet{2022Borchert_rise}, whose authors related these to outbursts of lower accretion rate, but feeding from a reservoir of material at a larger rotation radius, on the order of 40-50 AU. 

It is worth considering however, the chance that these stars may not be eruptive at all, and are instead exhibiting behaviour that is analogous to long duration `dippers' such as RW Aur \citep[see][]{2016Bozhinova_RW_Aur}. These long duration variables are traditionally detected from the fading event (which can take up to a couple of years to reach minima), rather than the return to the star's normal baseline luminosity. These events typically result in the star becoming redder, and thus they would demonstrate blueing when increasing in brightness, as we might be seeing for our long-duration rising YSOs \citep[for more information on this see][and references therein]{2013AA_TAU_Bouvier,2021Covey_AA_Tau}. This behaviour is not entirely the case for UX Ori type fades, which are likely
caused by re-emission of stellar radiation from disk scattering, and thus show both reddening and blueing events when fading \citep[see][for more details]{1999Herbst}. We still prefer the eruptive classification for these sources owing to the higher observed amplitudes, and longer durations than the previously observed `dippers'. 

\subsubsection{Intermediate Duration Eruptive YSOs}

\begin{figure}
    \includegraphics[width=0.95\columnwidth]{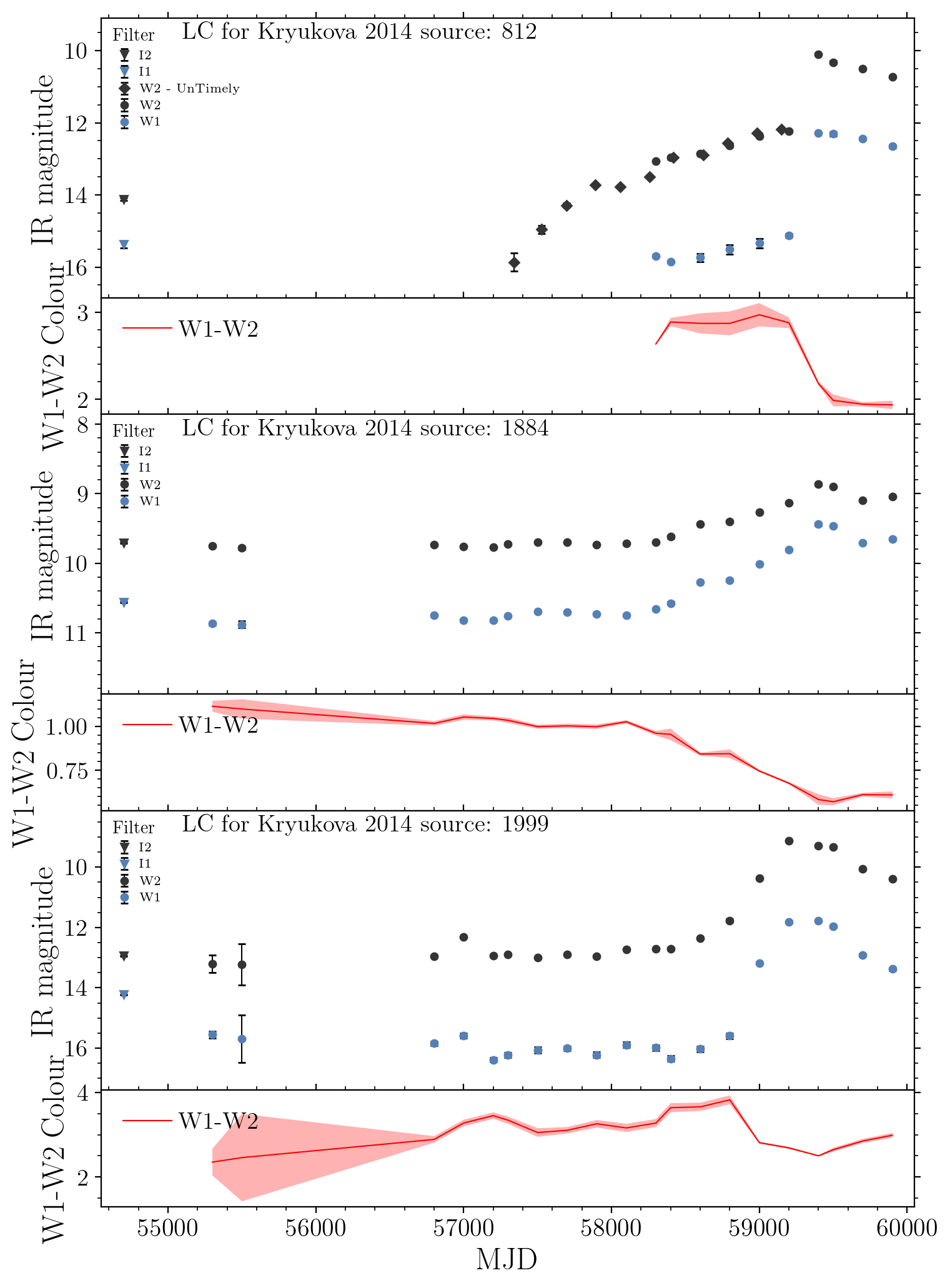}	
    \caption{NEOWISE $W1$ \& $W2$ light curves for three of the candidate eruptive variables with shorter duration, FUor-like outbursts. Each plot also contains a $W1-W2$ colour indicator (red line), with the error included as the pink-shaded region. Source 812 also contains W2 data from the UnTimely catalogue {\citep{2023unTimely}}.}
    \label{fig:FUors_kryu_IT}
\end{figure}

We identify eight stars in this sample with outbursts that are within the 2 to 10 yr boundary we set for the `Intermediate-Duration Eruptive' (IDE) group. Of these, three targets (Sources 812, 1884 and 1999) have morphologies that resemble a FUor-type event (see Figure \ref{fig:FUors_kryu_IT}), but are shorter than the >20-year timescales expected for FUors and are similar in morphology (if not amplitude) to VVV1636-4744 and L222\_4 from \citet{2024Zhen_FU}. The two stars from Guo et al's sample have similar rise times ($\sim$500 days) \& NIR amplitudes of 4 mag and 6 mag respectively, but have differing spectral characteristics. L222\_4 presents a typical red, absorption spectrum, typical of FUors, whereas VVV1636-4744 has an emission line dominated spectrum, more like EX Lup. Two further IDEs have completed outbursts (Sources 397 and 1017, Figure \ref{fig:YSOs_kryu_IT}). The final three IDEs (Sources 625, 658 \& 1562, in Figure \ref{fig:YSOs_kryu_IT_busy}) have a multi-peaked shape to their light curves that do not resemble those of FUors. 

Of the stars that resemble shorter duration FUors, we found some key differences between each star. For example, Source 812 displays at least 5 yr of a gradual increase in luminosity, before a very short (\textless 6 months) outburst. This behaviour is similar to FUors like Gaia17bpi (with the 2-stage initial outburst) as discussed earlier. Source 1999 has the second largest outburst in the sample but displays a much steeper decay slope than the three stars mentioned previously, indicative of a shorter outburst duration than most classical FUors, but longer than classical EXors (estimated at $\approx$8 years). Finally, Source 1884 has the second lowest amplitude of the stars in this group at 1.44 mag (with a gradient of $\approx0.36~\mathrm{mag~yr}^{-1}$), but this is combined with the bluest quiescent $W1~-~W2$ colour (between 1.11 mag and 0.56 mag). Given the range in spectra seen in \citet{2024Zhen_FU}, acquiring NIR spectra for these three sources is essential in ascertaining the nature of the outbursts we have observed.

\subsubsection{YSOs with Short-Duration Outbursts}

\begin{figure}
    \includegraphics[width=0.95\columnwidth]{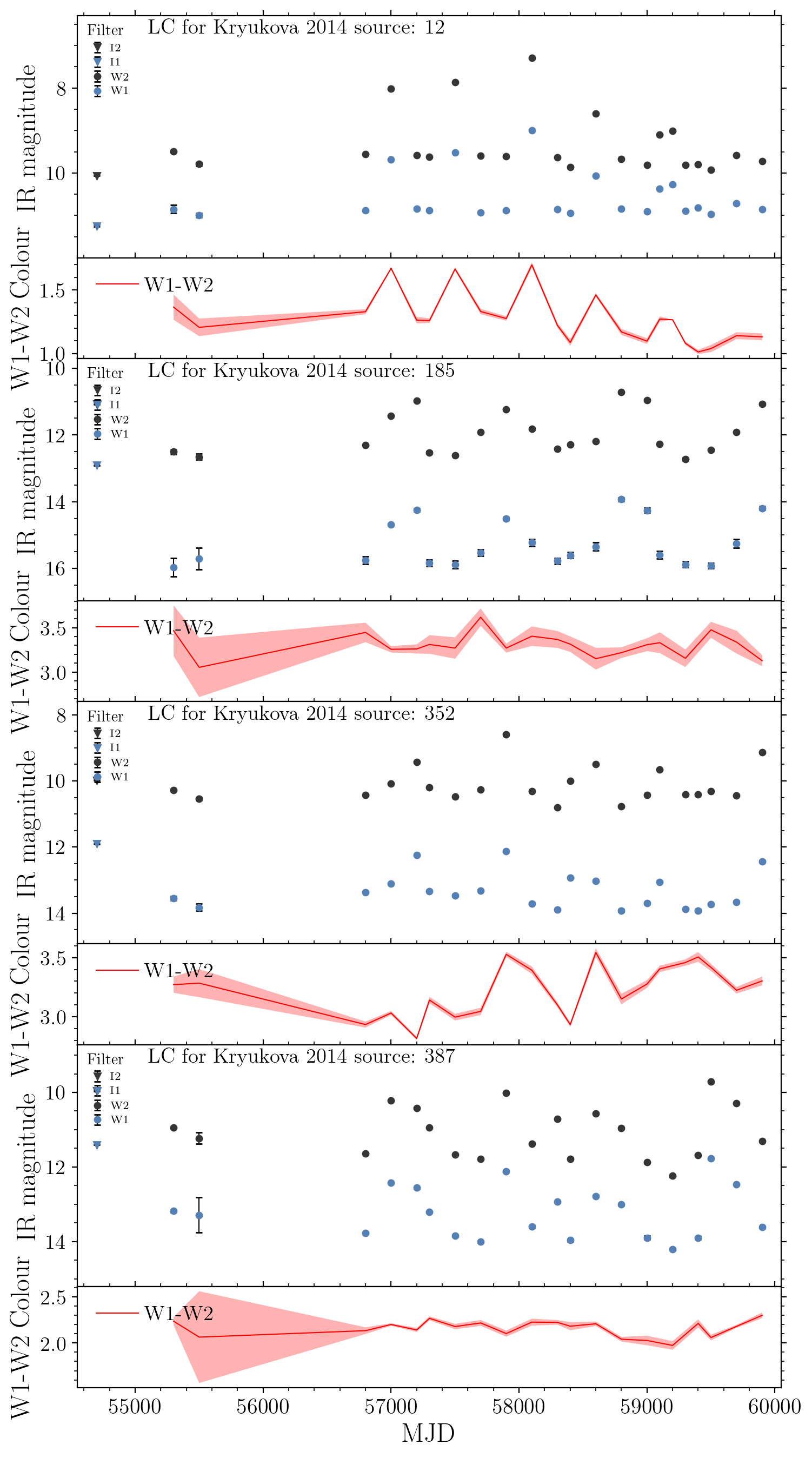}	
    \caption{NEOWISE $W1$ \& $W2$ light curves for the four semi-regular repeating short duration outbursts.}
    \label{fig:EXors_Kryu_Rep}
\end{figure}

\begin{figure}
    \includegraphics[width=0.95\columnwidth]{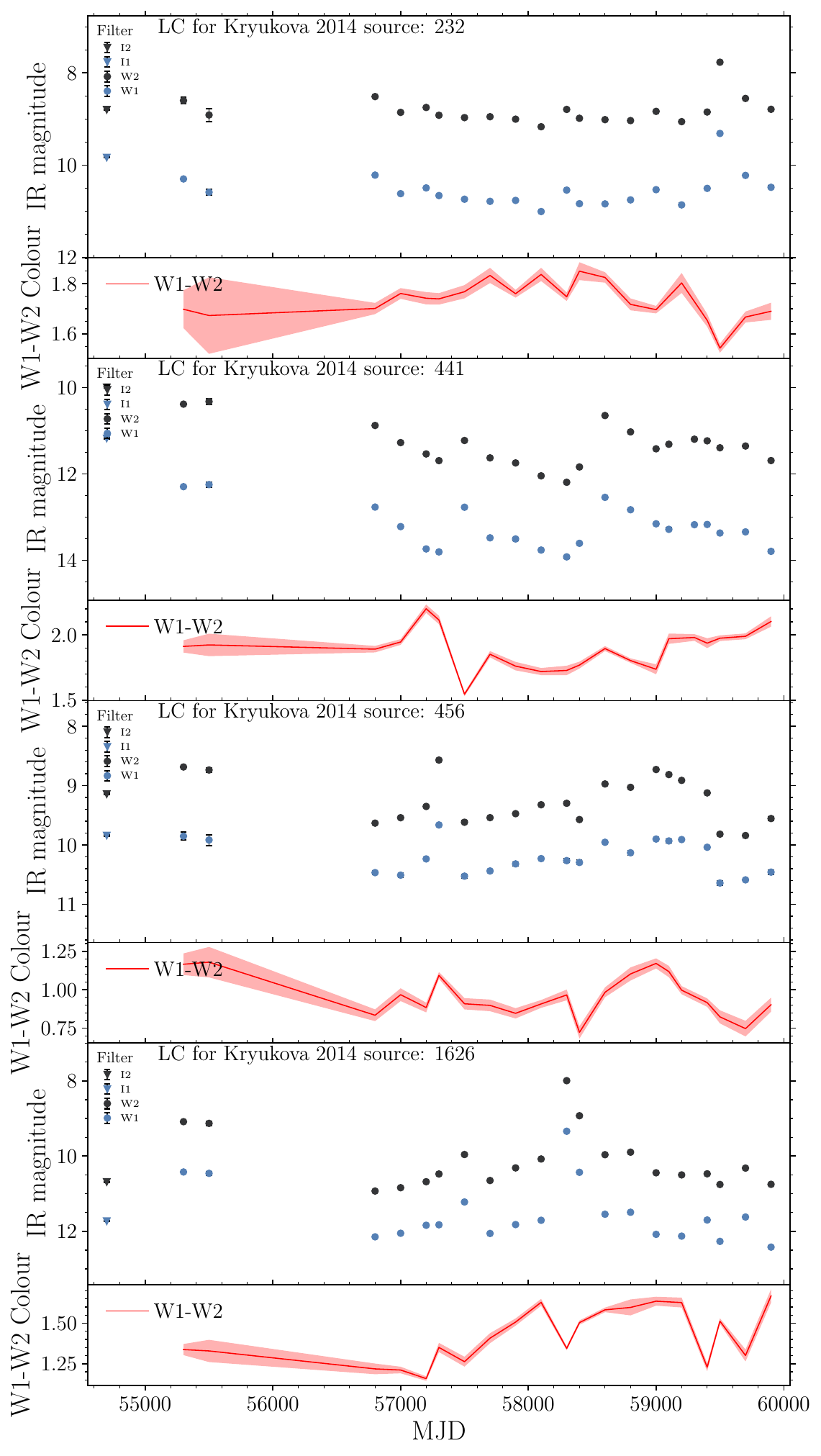}	
    \caption{NEOWISE $W1$ \& $W2$ light curves for four of the short-duration outbursts that have only one or two eruptive events.}
    \label{fig:EXors_Kryu}
\end{figure}

In total, there are 11 sources in this category, with the duration of outbursts less than two years. Among them, seven stars have multiple bursts (four of these are semi-regular, see Figure \ref{fig:EXors_Kryu_Rep}) and four others only have a single burst. The rest of the light curves are shown in Figures \ref{fig:EXors_Kryu} \& \ref{fig:Kryu_STEs_App}. There are no clear trends within this group, with a broad spread in amplitude, burst duration and colour behaviour. There are several sources with multiple outburst-like events of quasi-periodic or aperiodic nature in this group, which are of a more uncertain classification. It could be assumed that occultation from circumstellar material (as in many `dippers') would be periodic, and thus rule out the option for our repeating YSOs, but variation from circumbinary disks \citep[as in KH 15D][]{2016Aru_KH15D}, as well as from mis-aligned or warped disk components \citep[see][]{2017Rodri_Warps,2019Davies_Warps}, can deliver aperiodic extinction driven variation in YSOs. In the optical regime, a large proportion of dippers found are aperiodic \citep[see][]{2022Capistrant_dippers}, although the stars in those authors' samples were largely of a lower amplitude, and bluer colour than ours, and had access to a much shorter time baseline that NEOWISE is not sensitive to. On the whole the morphologies of our repeating events (largely symmetrical) more closely resemble the short term accretion-driven variations seen on many YSOs in the NIR, albeit at a higher amplitude. Similar behaviour was identified for a small number of YSOs in \citet{2018Wolk_YSOVAR}, which were also MIR detected sources, although their YSOs were of lower amplitude, and displayed no colour change over time.

Some sources of note in this selection are discussed below: 
\begin{itemize}
    \item Source 12 has four aperiodic outbursts with $>1$ mag in $W2$-band amplitude. Unlike a typical EXor, Source 12 has redder-when-brighter colours, although given the large spectral index (1.03) this could be a result of reprocessing of more of the accretion luminosity into the MIR. 
    \item Source 185 has four quasi-periodic outbursts that each last for $\sim500$ days (one of which is ongoing). We do not perceive the star to be a 'dipper' because of the irregularity of the bursts (which could not be reliably fitted by normal period-finding routines), which would normally - although not guaranteed to - be periodic if caused by extinction from orbiting material. There is a chance that this object might represent contamination from an evolved star (such as a long-period variable), although these stars' MIR light curves are usually more regular than seen here.  
    \item Source 232 has a recent outburst, in addition to the one previously identified in the UGPS survey \citep[included within][]{Lucas2017}. We will discuss it in more detail in \S\ref{sssec:232}. 
    \item Source 333 is of note from an additional long-duration outburst of 1.55 mag in $W1$, which has been ongoing since April 2019, although it reached its peak after $\sim 600$ days. This burst was accompanied by a small reduction in the $W1-W2$ colour, which was 2.13 at the start of the burst but fell to 1.50 at the maxima, and it continued to fall after the photometric maxima. 
    \item Source 1626 (Fig \ref{fig:EXors_Kryu}, bottom panel) has the largest single outburst amongst 
    the short-duration sources at 2.37 mag in $W1$, which combines with a slight blueing $W1-W2$ colour of 0.28. The single large burst lasts only for $\sim 500$ days, all of which are indicative of an EXor-type eruption. 

\end{itemize}
\subsubsection{Source 232}\label{sssec:232}

\begin{figure}
    \includegraphics[width=0.98\columnwidth]{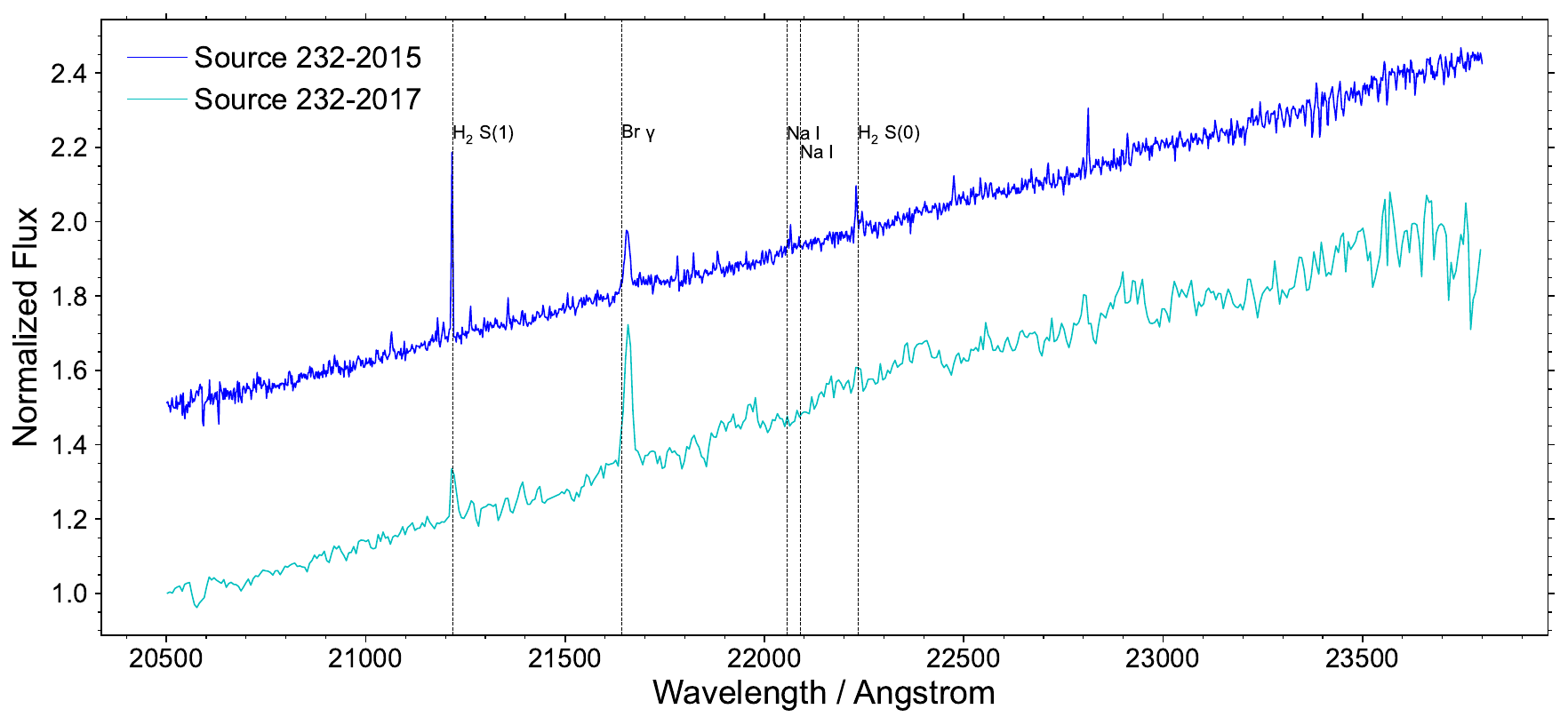}	
    \caption{$K$ and $K'$ bandpass spectra for Source 232, taken in 2015 and 2017. Note the absence of lines from sodium in the 2017 spectrum, normally seen as a tracer of the accretion stream.}
    \label{fig:232_spec}
\end{figure}
Source 232, also known as GPSV28 \citep{2014CarlosCP} and IRAS 20226+4206, was observed to have probable outburst in December 2008 with $\Delta K_{s} = 1.0$ mag, originally detected by the UGPS survey (included as Source 463 in \citet{Lucas2017}), the amplitude increases to $\Delta Ks = 1.56$ mag when including 2MASS photometry. Additionally, it has two spectroscopic follow-up observations using Gemini/NIFS (2015) and Subaru/IRCS (2017). 
These spectra are shown in Figure \ref{fig:232_spec}, which were acquired $\sim$4 and $\sim$6 years after the previous outburst. The spectra show notable differences in the emission line ratios, wherein the 2015 spectrum has more prominent $\mathrm{H}_{2}$ lines (S(1) and S(0)) than \ion{H}{i}, and a clear \ion{Na}{i} doublet. The implication is that flux is a combination of the stellar photosphere, the inner accretion disk and outflows (traced by $\mathrm{H}_{2}$). By the 2017 the spectrum is dominated by the Br$\gamma$ emission line, with a 3.98$\pm0.61~\mathring{\mathrm{A}}$ (from $-1.98\pm0.58~\mathring{\mathrm{A}}$, to $-5.95\pm0.2~\mathring{\mathrm{A}}$) increase in equivalent width from the previous spectrum. Considering that this change implies an increase in the accretion rate, we speculate that a higher proportion of the observed flux is from the hot inner accretion disk, hence why emission lines associated with the accretion streams (\ion{Na}{i}) are now significantly veiled by the continuum. Whilst the initial outburst of Source 232 was not fully captured by the WISE light curves, a second burst ($\Delta W1$ = 1.4 mag) was observed between 2021 and 2022, making clear that this star has EXor-like behaviour. 

\subsubsection{Eruptive YSOs with Ambiguous Outburst Durations}

The final three stars in this sample have ongoing outbursts, that are yet to reach the plateau, making it a challenge to determine a likely outburst duration (See Figure \ref{Fig:Ambig_YSOs} for light curves). Source 2003 is characterised by a large amplitude, 5.1 mag in W1, and colour change of over 1 mag to the blue in $W1-W2$  during the burst, both of which are common among FUor-type stars. With the outburst being unfinished, there is a possibility of it being one of the largest MIR outbursts yet detected for a FUor. 

The remaining two stars in the group, sources 1738 \& 1769, display similar long-term trends. Each star's light curve contains a small, short-term increase in brightness, before a visibly longer-term outburst (ongoing). No other stars in the sample show such activity, and as yet the nature of the current outbursts is unknown.  
\begin{table*}
    \centering
    \label{tab:SPICY_EVs}
    \caption{35 EV candidates found in SPICY with amplitude $>$1 mag}
    \begin{tabular}{lcccccc} 
    \hline
    \hline
    SPICY ID & RA (deg) & Dec (deg) & $\Delta W1$ & $(W1 - W2)_{\rm min}$  & Colour Change & SED Class \\
 & J2000 & J2000 & mag & mag  & (when brighter)  &  \\
    \hline
        107699  & 304.0523 & +39.0873 &  1.025 $\pm$ 0.010  &  0.321  & Redder & Class II\\ 
        107742  & 304.1969 & +39.3892 &  1.054 $\pm$ 0.010  &  1.338  & Bluer & Class II\\ 
        107855  & 304.3484 & +39.3052 &  0.969 $\pm$ 0.008  &  0.550  & No Change & Class II\\

        108504  & 305.3081 & +38.9968 &  1.125 $\pm$ 0.013  &  0.441  & Redder & Class II\\ 
        108553  & 305.3903 & +40.8526 &  0.855 $\pm$ 0.017  &  0.698  & Redder & Class II\\ 
        108835$^{\ddagger}$  & 305.7091 & +40.5112 &  1.338 $\pm$ 0.046  &  0.786  & Bluer & Class II\\
        109025  & 305.9034 & +39.5591 &  0.875 $\pm$ 0.006  &  0.103  & Redder & Class III\\

        109259$^{\ddagger}$  & 306.1053 & +42.4036 &  1.233 $\pm$ 0.014  &  1.091  & Bluer & Class II\\ 
        109392  & 306.1617 & +38.5000 &  1.904 $\pm$ 0.015  &  0.332  & Redder & Class II\\
        109705  & 306.4289 & +38.5806 &  4.121 $\pm$ 0.021  &  1.171  & Bluer & FS\\
        109973  & 306.6403 & +39.6054 &  0.817 $\pm$ 0.021  &  0.345  & Redder & Class II\\ 
        110521$^{\ddagger}$  & 307.0781 & +39.4852 &  0.836 $\pm$ 0.007  &  0.157  & Redder & Class II\\ 
        110601  & 307.1598 & +38.8962 &  0.931 $\pm$ 0.010  &  0.416  & Redder & Class III\\ 
        110991  & 307.4678 & +40.4054 &  1.315 $\pm$ 0.008  &  0.713  & Redder & Class II\\ 
        111048  & 307.5187 & +40.3839 &  1.677 $\pm$ 0.011  &  0.025  & Redder & Class II\\ 
        111336$^{\ddagger}$  & 307.7453 & +40.5380 &  1.522 $\pm$ 0.036  &  -0.124  & Redder & Class II\\ 
        111413$^{\ddagger}$  & 307.7927 & +40.4144 &  1.165 $\pm$ 0.023  &  0.443  & Redder & FS\\ 
        111739$^{\ddagger}$  & 307.9553 & +38.9405 &  1.810 $\pm$ 0.067  &  1.346  & Bluer & ?\\ 
        111836  & 307.9984 & +41.0660 &  1.125 $\pm$ 0.014  &  0.821  & Redder & Class II\\ 
        111892  & 308.0221 & +42.8133 &  0.876 $\pm$ 0.008  &  1.050  & Varied & FS\\
        112458  & 308.2766 & +40.7189 &  0.616 $\pm$ 0.057  &  -0.420  & Redder & Class II\\ 
        112533  & 308.3147 & +41.0470 &  1.561 $\pm$ 0.018  &  0.007  & Redder & Class II\\ 
        112884  & 308.4971 & +41.4482 &  0.614 $\pm$ 0.011  &  0.180  & Redder & Class II\\ 
        112979$^{\ddagger}$  & 308.5504 & +40.7098 &  1.201 $\pm$ 0.006  &  0.442  & Redder & Class II\\ 
        113094$^{\ddagger}$  & 308.6131 & +42.2315 &  1.672 $\pm$ 0.001  &  0.902  & Bluer & Class II\\ 
        113145  & 308.6433 & +41.1734 &  1.489 $\pm$ 0.015  &  0.643  & Redder & Class II\\ 
        113564  & 308.9150 & +41.8482 &  1.550 $\pm$ 0.016  &  0.609  & Varied & Class II\\ 
        113803$^{\ddagger}$  & 309.0589 & +41.1173 &  0.904 $\pm$ 0.009  &  0.166  & Redder & Class III\\ 
        113865$^{\ddagger}$  & 309.0924 & +42.4434 &  1.043 $\pm$ 0.049  &  -0.100  & Varied & FS\\ 
        113929  & 309.1289 & +39.6458 &  0.940 $\pm$ 0.005  &  0.441  & Redder & Class II \\ 
        114037  & 309.1980 & +42.6153 &  1.001 $\pm$ 0.051  &  -0.018  & Varied & Class II\\ 
        114471$^{\ddagger}$  & 309.5422 & +42.2826 &  1.371 $\pm$ 0.015  &  0.417  & Bluer & Class II\\
        114590  & 309.6542 & +42.5540 &  1.603 $\pm$ 0.034  &  0.542  & Redder & FS\\
        115546  & 310.4321 & +39.6725 &  0.662 $\pm$ 0.023  &  0.732  & Redder & FS\\ 
        115613$^{\ddagger}$  & 310.5619 & +39.7890 &  1.158 $\pm$ 0.010  &  0.777  & No Change & FS\\ 
    \hline
    \hline
    \end{tabular}
    \flushleft{The colour behaviour column notes how the $W1-W2$ colour changes during the eruptive portions of the light curve: Bluer means the source is bluer-when-brighter (much like traditional EXors), Redder indicates redder-when-brighter behaviour, and "Varied" notes stars whose colour behaviour changes over the total duration of the light curve. YSOs with $W1$ amplitudes under 1 mag have $\Delta W2 > 1$ mag. Sources marked with $^{\ddagger}$ indicate that the YSO has a light curve morphology that is hard to be certain of an accretion driven outburst.}

\end{table*}

\subsection{Cygnus-X YSOs without [24] detections, from SPICY}\label{ssec:spicy_sources}

Among the sample of the SPICY members,  402 stars have amplitudes of $>1$ magnitude in either $W1$ or $W2$-band. We chose 1 mag as the threshold for inspection because MIR amplitudes are often a factor of 2 lower in amplitude than those seen in optical band-passes, and as such this corresponds to $\approx$2 mag, a little lower than the guidelines for an outburst as laid out by \citet{2022_PP7_var}. The majority of the 402 stars showed single-epoch dipping events (likely extinction induced) or repeating redder-when-brighter 'dipper' behaviour. We then manually selected 35 EV candidates (see Table \ref{tab:SPICY_EVs}), including seven FS sources, 24 Class II YSOs, and three Class III YSOs. An additional source is considered uncertain (it lacks photometry in multiple \textit{Spitzer} bandpasses), although its average WISE colour ($W1-W2 = 1.66$) is comparable to the bluer Class I sources from section \ref{sssec:E_FU}.

Across the sample of EVs, 23/35 stars show the redder-when-brighter trend in their $W1-W2$ colour during outburst. Applying the outburst duration-based classifications defined in \S\ref{sec:results}, we find 30 SDEs, one IDE, two LDEs, and two sources of ambiguous classification.

\subsubsection{New FUor-type/Long Duration Eruptive YSOs}

\begin{figure}
    \includegraphics[width=0.98\columnwidth]{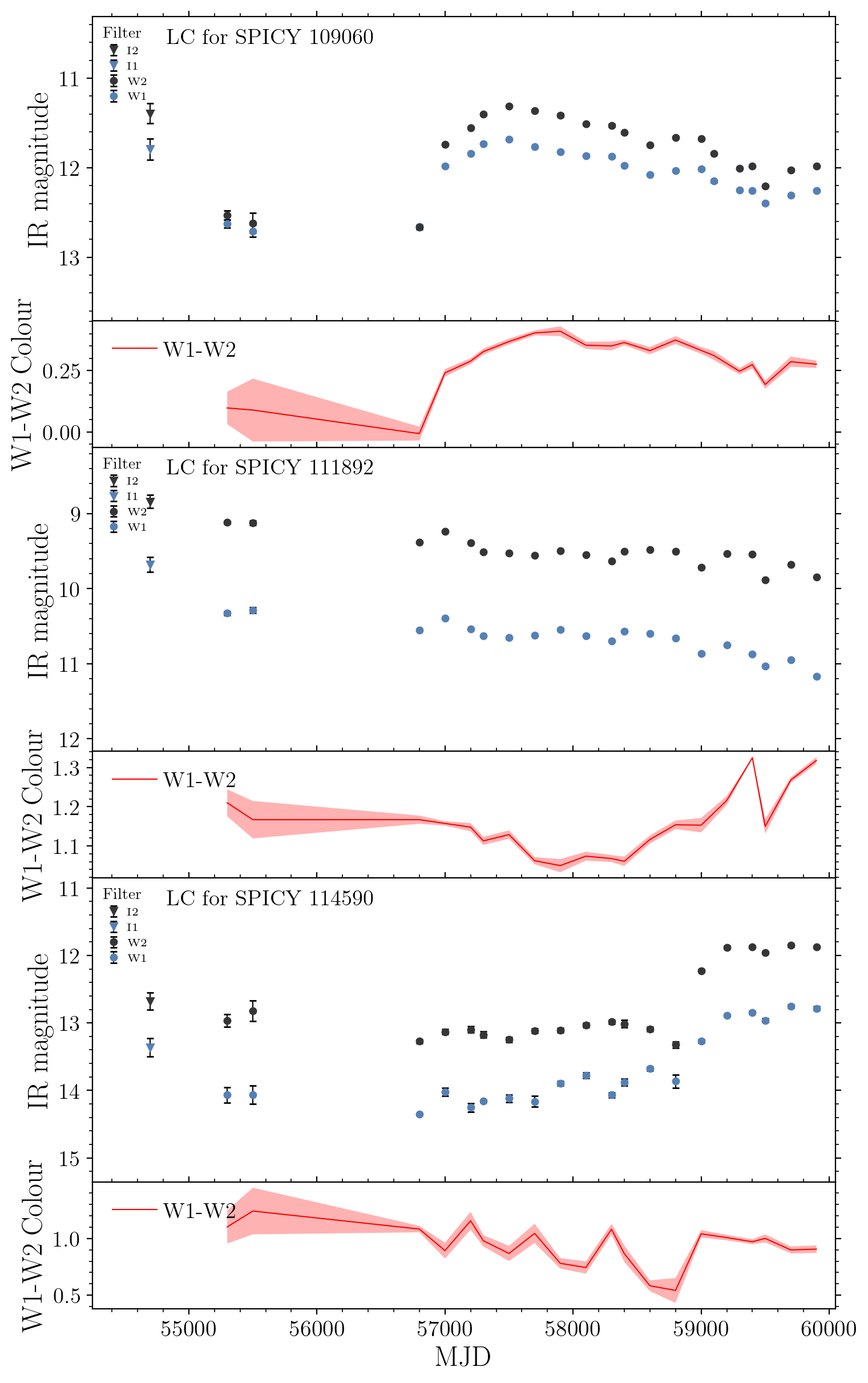}	
    \caption{NEOWISE $W1$ \& $W2$ light curves for three FUor candidates from the SPICY selected Class II/Flat-spectrum sources. }
    \label{fig:FUors_spicy}
\end{figure}

The outburst detection code described in \S\ref{sec:methods} located one star (SPICY 114590) with a shape comparable to that of an archetypal FUor (although the rise time is still on the longer scale, more similar to the VVV identified FUors), although it has a peak amplitude of $< 2$ mag (see Figure \ref{fig:FUors_spicy}). The MIR colour behaviour of SPICY 114590 is different to that seen in the FUor candidates in \S\ref{ssec:embedded_sources}, as this object gets redder during the outburst. The low amplitude is not unexpected in the general sense, as it can be predicted by outburst models for YSOs with already enhanced accretion rates \citep[see Figure 1 in][with the caveat that this was for a sun-like star]{2022Hillenbrand_FUorAmp}.

SPICY 111892 has been previously identified as a candidate EV in UGPS by \citet{2015PhDCarlos} and \citet[][as GPSV64]{Lucas2017}. It displayed a 1.5 mag increase in $K_s$-band brightness between 2006 and 2009, which would be low for a FUor if this is the full amplitude of the event. However, the non-detection in earlier 2MASS images implies a longer rise time and a higher amplitude.

The lower detection limit of the 2MASS $K_s$ image in which SPICY 111892 resides is calculated at 16.0 mag. This was calculated by running the \texttt{photutils} method called \texttt{ImageDepth}. This method takes the zero-point for the instrument, and a selection of magnitudes of stars in the image, and computes the lowest detectable flux for a given precision in sigma (quoted here are 3$\sigma$ values) for a large number of randomly placed apertures. 

\begin{figure}
    \includegraphics[width=0.98\columnwidth]{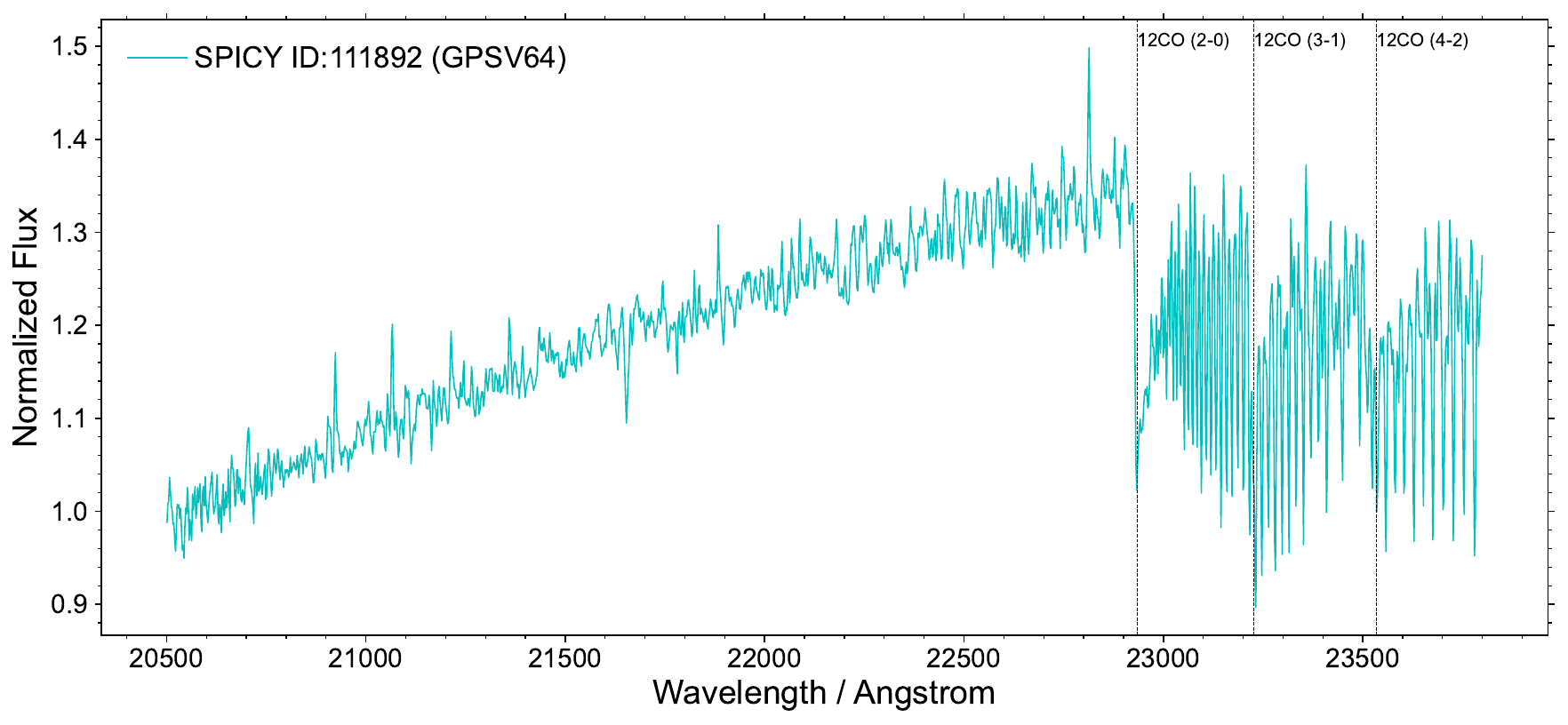}	
    \caption{Gemini/NIFS spectrum for SPICY 111892. The spectrum is FUor-type, with a featureless red continuum and strong $^{12}$CO absorption features.}
    \label{fig:S500_Spec}
\end{figure}

Thus the total $\Delta K_s$ has a lower limit of 4.0 mag, with a rising timescale of between $\sim$1000 and  $\sim$4000 days. A follow-up spectrum was obtained in December 2014 (Figure \ref{fig:S500_Spec}) with Gemini/NIFS, which shows the mostly featureless red continuum and CO bandheads often associated with FUor spectra (the emission line at 2.282$\mu$m is likely spurious, as it had appeared in a large number of other YSOs in the same set of observations). See further examples of FUor-type spectra in \citet{Connelley2018} and \citet{2024Zhen_FU}.

\begin{figure}
    \includegraphics[width=0.98\columnwidth]{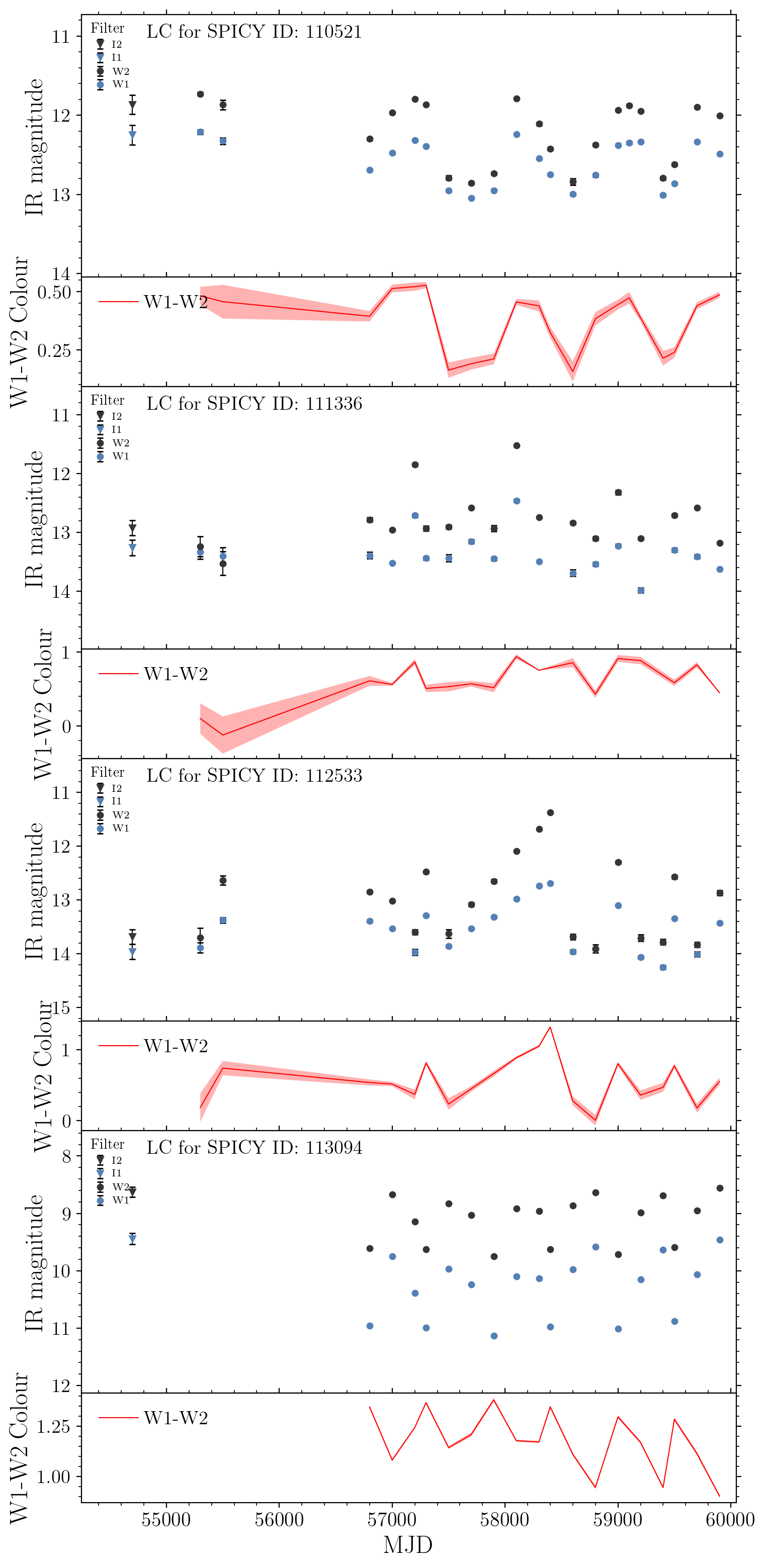}	
    \caption{NEOWISE $W1$ \& $W2$ light curves for four EXor candidate eruptive variables. These are selected because they demonstrate a wide range of behaviours and outburst durations. Of particular note is Source 112533, which had periodic small outbursts after a much larger, long-duration event.}
    \label{fig:EXors_main_spicy}
\end{figure}
\subsubsection{New EXor-like/Short Duration Eruptive YSOs}

Of the 35 EV candidates in this group, 86\% (30) have short-duration outburst events, many of which are similar to EXor-like YSOs. In line with the known diversity of these systems, we observed significant variation in the light curve morphologies, with differences in burst duration, amplitude, colour and repetition. We present four examples in Figure \ref{fig:EXors_main_spicy} that illustrate the variety (the rest can be seen in Appendix \ref{app:EXors}). 

SPICY 113094 displays quasi-periodic behaviour at a roughly 550 days period\footnote{fitted using the \texttt{astropy LombScargle} function.}, with each burst of up to 1.6 mag and bluer colours during the photometric maxima. This is characteristic of stars such as SPICY 116663 \citep{2021Kuhn_spicy} and periodic outbursting candidates discovered in VVV \citep{2022Zhen_periodic}, 
albeit with a longer period than most examples.

SPICY 110521 shows repeating burst behaviour, although without any clear periodicity, effectively ruling out periodic occultation by dust. 
With outbursts of over 1 mag and colours that redden with increases in luminosity, it is however unlike the typical behaviour of EXors \citet[see][]{Lorenzetti2012}, wherein colours are often expected to become bluer during outbursts. The behaviour seen in SPICY 110521 is commonly seen in this sample: 20 of 33 (60.6\%) short duration (EXor-like) eruptive sources are redder in outburst. 

SPICY 111336 also displays aperiodic, repetitive outbursts, but these are of differing amplitude, in contrast to many of the other repeating sources. The amplitudes in $\Delta W2$ range from $\sim 0.7$ mag to $\sim 1.5$ mag, although they are of a similar $\sim 400$ day duration (the duration is an upper limit owing to the wide sampling of NEOWISE). The range of amplitudes could also be a result of the sampling, with the peak of the outbursts being missed at several epochs, especially if the true period is less than 400 days.

The final EXor-like star to be discussed will be SPICY 112533, which is the most unusual star in this sample of later-stage YSOs. Seemingly there are two components to the light curve, with a potential long-term fading trend, which is separated into five detected outbursts each of between $\sim 0.8~\&~1.0$ mag and $\sim 1.0~\&~2.0$ mag in $W1~\&~W2$ respectively. Of additional note is the single outburst of at least 900 days (between 2016 and 2018) which bears little resemblance to other EXor-like bursts, owing to the comparatively short decay timescale as compared to the rise time.    

Overall the sample of short-term variables is slightly biased towards sources with repetitive outburst behaviour, as is the case for 60.6\% (20/33) of the sample. Whilst none of these are truly periodic, five of the stars display quasi-periodic bursts, with periods of over 400 days (the lowest that can be reliably identified with NEOWISE given the sampling).  

\begin{figure}
    \includegraphics[width=0.98\columnwidth]{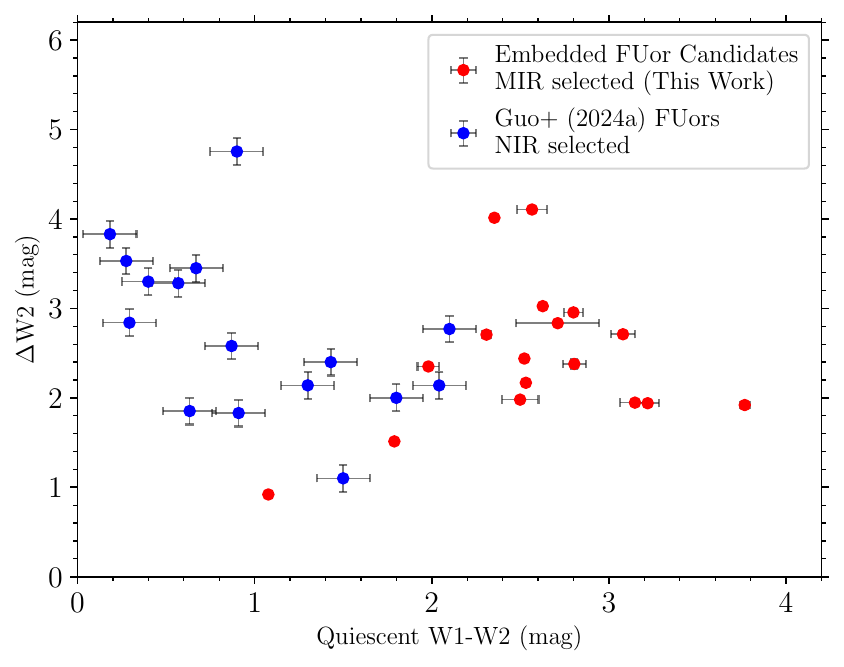}	
    \caption{$W2$-band amplitude against quiescent $W1$-$W2$ colour for EVs discussed in \S\ref{ssec:embedded_sources}. The NIR selected FUors are adopted from \citet{2024Zhen_FU}.}
    \label{fig:W2vColour}

\end{figure}
\begin{figure*}
    \includegraphics[width=\columnwidth]{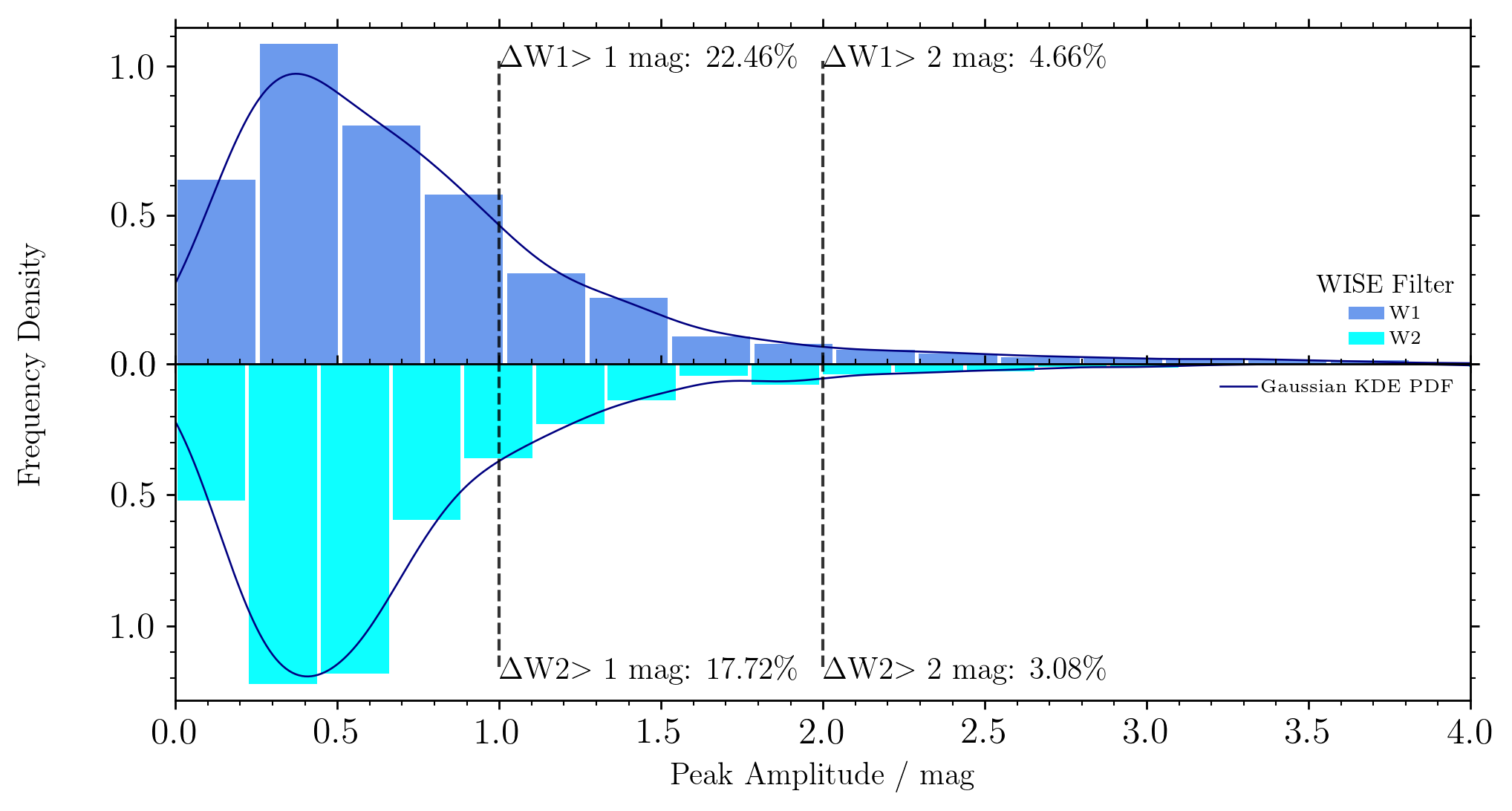}
    \includegraphics[width=\columnwidth]{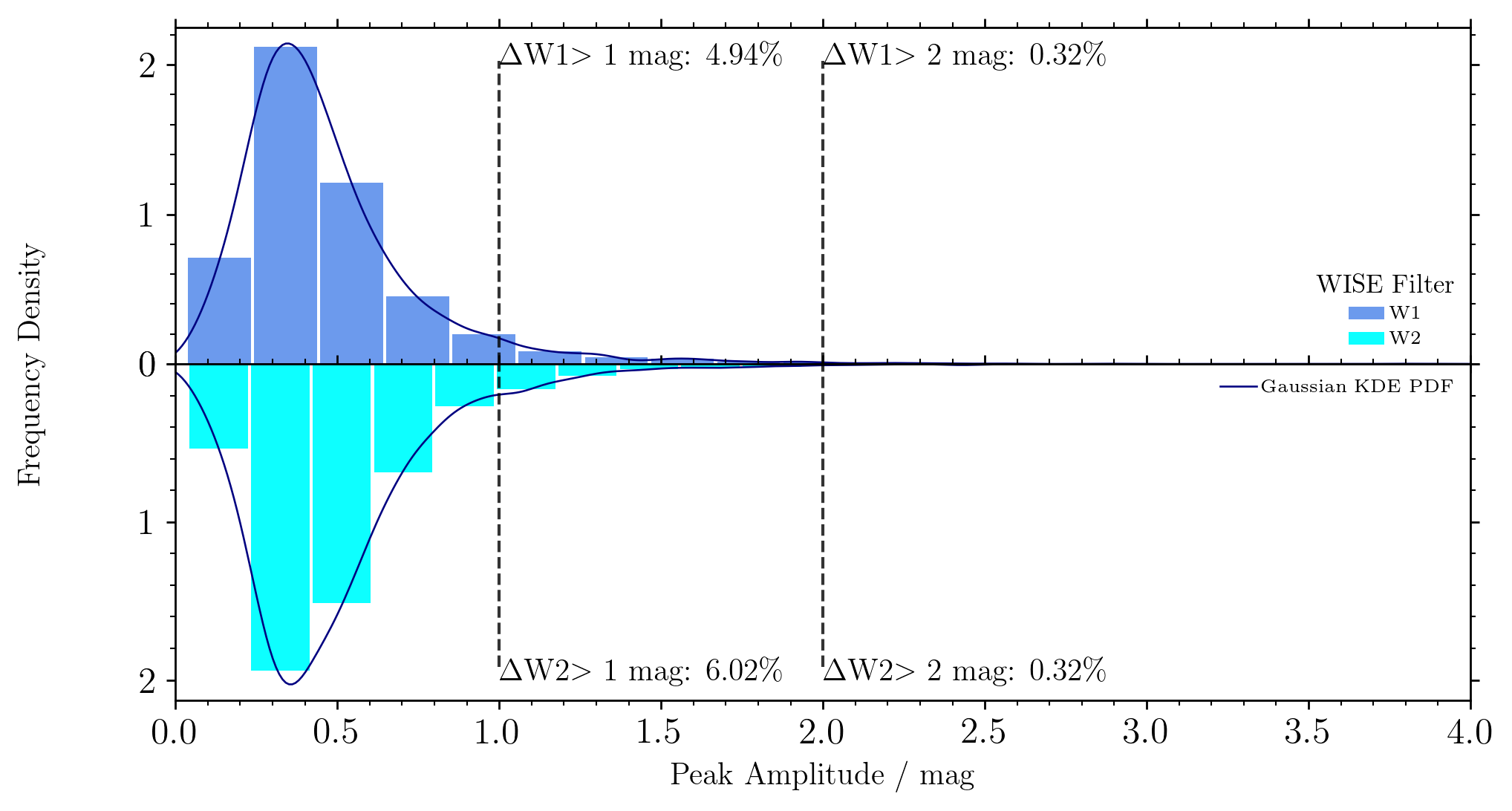}	
    \caption{Histograms displaying peak amplitudes in {$W1$ \& $W2$} for both of the two samples discussed in Section {\ref{sec:results}}. The left plot is the $[24]$ detected sample, and the SPICY selected C2 and FS sources are in the right panel. Each individual histogram is presented with a probability distribution function, created by performing a Gaussian kernel density estimation.}
    \label{fig:AmplitudeHistComparison}
\end{figure*}
\section{Discussion}\label{sec:discussions}
\begin{table}
	\centering
	\label{tab:FUors}
 \caption{List of the FUor candidates from this work, both samples and all durations.}
	\begin{tabular}{lccc} 
 \hline
        Source ID & Timescale & Amplitude & Colour Change \\
        (this work) & Category & ($W1$) & (when brighter)\\
        \hline
        \multicolumn{4}{|c|}{Likely Candidates} \\    
        \hline
        257  & LDE & 3.28 & Bluer\\ 
        294  & LDE & 2.52 & Bluer \\ 
        362  & LDE & -$^{*}$ & Bluer \\ 
        591  & LDE & 2.35 & Bluer\\ 
        812  & IDE & 3.57 & Bluer\\
        1048 & LDE & 3.98 & Bluer\\
        1475 & LDE & 2.81 & Varied\\ 
        1884 & IDE & 1.44 & Bluer\\
        1945 & LDE & 1.96 & Bluer\\
        1964 & LDE & -$^{*}$ & Redder\\
        1991 & LDE & 2.82 & Bluer\\
        1999 & IDE & 2.35 & Bluer\\
        2003 & Amb & 5.09 & Bluer\\ 
        111892 & LDE & -$^{*}$ & Varied\\
		\hline
        \multicolumn{4}{|c|}{Potential Candidates} \\ 
        \hline
        397  & IDE & 2.26 & Bluer\\
        880  & LDE & 2.63 & Varied\\
        912  & LDE & 2.86 & Bluer\\
        1017 & IDE & 2.31 & Redder\\
        1738 & Amb & 2.40 & Bluer\\
        114590 & LDE & 1.60 & Redder\\
  \hline
	\end{tabular}
\flushleft{$^{*}$: These stars had their outbursts discovered with NIR data, so the $W1$ amplitudes are not included. Also note that the stars labelled as IDEs have their timescales estimated via their initial fading period, and could still take longer than 10 years to return to quiescence.}

\end{table}    

\subsection{Comparison of Embedded FUor Candidates with Classical FUors}
Given the comparatively low number of known FUor-like stars, our discovery of 13 (these being the stars with FUor-type light curve morphologies or spectra) potential new candidates within a single SFR warrants additional investigation. Given that our main sample was of the most embedded EVs within Cygnus-X, comparing their MIR behaviour to that of FUors discovered via a `traditional' optical outburst could reveal differences between the two populations. 

\citet{2024Zhen_FU} provided a selection of FUors discovered through analysis of the VVV near-infrared time series and further confirmed by follow-up spectroscopic observation. They found a negative correlation between the $\Delta W2$  and quiescent $W1 - W2$ colour (Fig \ref{fig:W2vColour} - blue points), whereas this does not agree with the long-duration EVs in our sample (Fig \ref{fig:W2vColour} - red points). The correlations were tested by performing Kendall $\tau$ tests on each sample, finding the NIR selected sources to have $\tau=-0.42$, compared to $\tau=0$ for the MIR selected EVs from this work.    
Our EVs from \citet{Kryukova2014} seem to have no correlation between $\Delta W2$ and $W1 - W2$,  although all still show the common bluer-when-brighter colour behaviour during the outburst, similar to FUors.
The figure also places the slowly rising eruptive variables amongst the FUor candidates, giving further weight to our hypothesis that these sources are YSOs that are building up to a FUor-type outburst (which may be optically detectable), in a similar vein to Gaia17bpi. 

The reasoning for the break in the correlation of the colour/amplitude relation for the embedded sources is not yet fully clear, although a thicker envelope (and thus redder quiescent colours) might imply that we are seeing greater reprocessing of the accretion luminosity into the MIR, inflating the observed amplitude.  

\subsection{Less-than-certain Short Duration YSO Outbursts}

Distinguishing short-duration eruptive events from dippers within our sample is an uncertain task, owing to the long cadence of the NEOWISE observation windows. The morphology of individual events can be hard to determine, and the colour behaviour during brightening can be impacted by numerous factors including inclination and the faint nature of some of our targets (see Section \ref{ssec:colours}). Previous work on the MIR variability of YSOs with the \textit{Spitzer} YSOVAR programme \citep{2004Werner_Spitz,2011MC_YSOVAR} focused on variability with timescales of weeks to a month (which would likely appear without a clear structure in NEOWISE). In \citet{2018Wolk_YSOVAR}, authors found an average amplitude of 0.21/0.13 mag in IRAC \citep{2004Fazio_Irac} $I1$ and 0.17/0.13 mag in $I2$ for Class I/II sources in the star-forming region Serpens South ($I1$ \& $I2$ are roughly analogous $W1$ \& $W2$). Notably, none of the YSOs in the Serpens South samples have an amplitude greater than 1 mag. In the Orion nebula, only 13 out of 2238 sources reached 1 mag in their I1 or I2 amplitudes on the timescale of two weeks. 

Our selection of stars with at least one bandpass with amplitudes of $>$1 mag was intended to remove the majority of stars with the type of short-term variance seen previously. It remains to be seen how successful this has been, with several (19) YSOs having light curve morphologies which are far removed from those expected of eruptive stars (see marked$^{\ddagger}$ YSOs in Tables \ref{tab:Embedded_EVs} \& \ref{tab:SPICY_EVs}). These stars not only carry the possibility of being dippers, but they could also belong to the loose grouping of 'Protostellar Outbursts (Infrared)', which was detailed in \citet{2022_PP7_var}. These shorter duration events of around 1-2.5 mag in the NIR, would correspond to $\sim$0.6-1.5 mag in the WISE bandpasses (Contreras Pe\~{n}a et al., in prep); a similar region to where many of our SDE's that do not resemble EXors (or repeating EXors) reside. Follow-up spectroscopy of these stars may shed further light onto the causes of the observed variability, and thus determine if they are from a single, cohesive group. Accounting for these less-than-certain eruptions, we can say that we have identified 48 strong candidate EVs, with a further 20 uncertain.

\subsection{Comparing the Rates of Eruptive Behaviour Between Embedded and Non-Embedded Sources}
\begin{figure}
    \includegraphics[width=0.98\columnwidth]{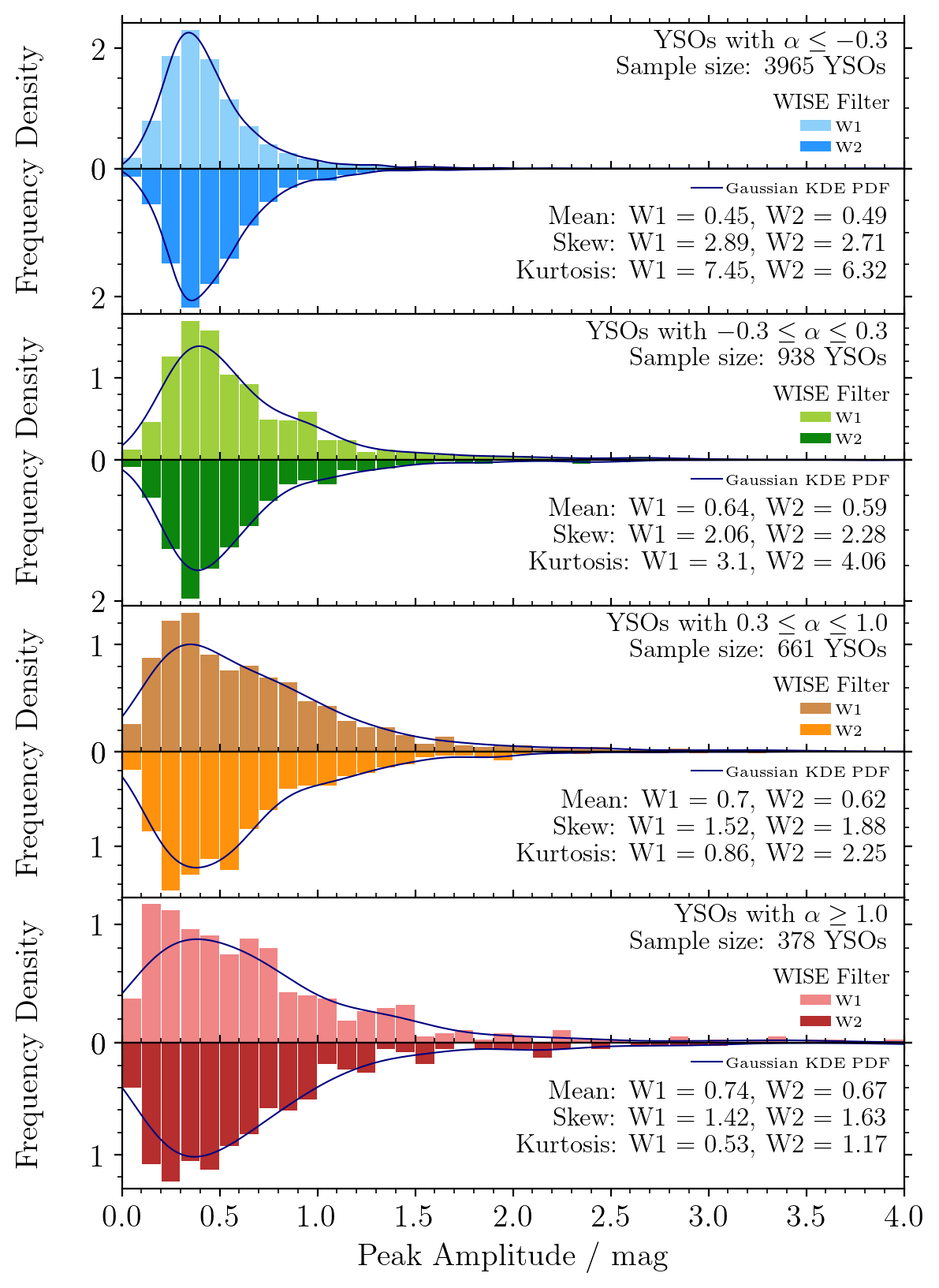}
    \caption{$W1$ and $W2$ amplitude histograms for the combined YSO sample, split by SED slope class ($\alpha$).}
  \label{fig:AlphaHists}
\end{figure}
\begin{figure}
    \includegraphics[width=0.9\columnwidth]{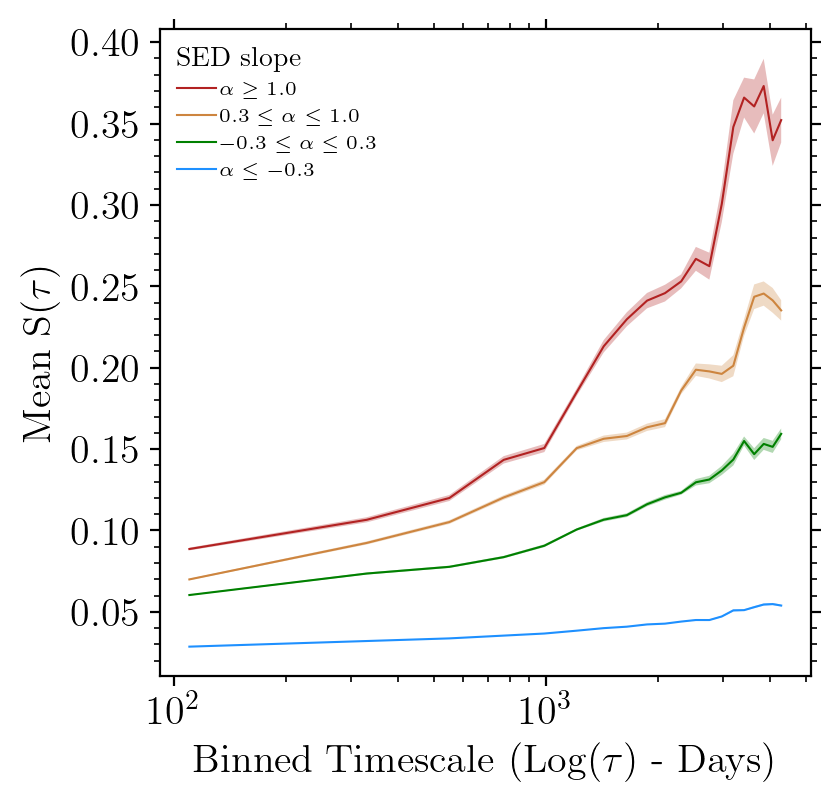}
    \caption{Structure functions across a range of timescales for the combined YSO sample. Each point is the median structure in each timescale bin, averaged for both $W1$ and $W2$. These are once again separated by the value of the $\alpha$ parameter for the SED slope}
  \label{fig:AlphaStrucs}
\end{figure}

By having a control sample of known YSOs, we can compare both the detection rates and amplitudes of eruptive variability for our embedded YSOs to those of the less embedded or more evolved sources within Cygnus-X. The comparisons between the frequency density and amplitude for our samples are shown in Figure \ref{fig:AmplitudeHistComparison}, with embedded Class I sources in the left panel, and FS/Class II sources in the right.
The embedded sources have a higher proportion of high-amplitude variables than the later stage sources, as 22.46\% of Class I YSOs (From the 24$\mu$m selected sample) have $\Delta W1 >1$ mag whilst just $4.94\%$ FS/Class II YSOs (the SPICY sample) reach the same amplitude. We note that only a proportion of these variables are eruptive, with large numbers of dippers, short-term variables and stars on long-term fading trends. Nevertheless, the fading sources may well be eruptive YSOs, wherein a large outburst has happened before our observation window and we are now witnessing the post-outburst cooling phase

To test the significance of these results, we performed a Mann-Whitney U (MW-U) test, between both samples. Because the sample of FS/Class II sources is substantially larger than that of Class I sources, 
we performed the MW-U test on two equal-sized samples drawn from distributions defined by a Gaussian kernel density estimation (fitted to all four histograms in Figure \ref{fig:AmplitudeHistComparison}). Both of these routines are carried out in \texttt{Python}, using the \texttt{gaussian\_kde} routine in \texttt{scipy.stats}. This test produced an average p-value of 3.46\% over 10000 iterations and thus provided moderately statistically significant evidence that these two samples were drawn from two different distributions. 

From the perspective of YSO evolution, the above result is within expectation, as younger systems have a larger reservoir of material to accrete from. Massive discs with envelopes are also more likely to become unstable under gravity, especially when considering that flyby events are likely to be more common in younger (less bound) clusters. For further discussion of this idea, see \citet{Audard2014}.  

Given the potentially higher chance for outbursts, searching for EVs amongst younger systems specifically should result in a larger sample, especially of the rarer FU Ori-type objects. This falls into the long-held view that longer duration outbursts happen in younger systems \citep{1996_Hart_Ken_FUors}, given the long average length of FUor events. The differing morphologies among our long-duration outbursts also fit into the ideas of \citet{2007Quanz}, which describe a possible evolutionary sequence for FUors, based upon their MIR features. The above point makes a strong case for follow-up spectroscopy at both near and mid-infrared wavelengths.

To further investigate the relationship between outburst amplitude and age for YSOs, we compared the combined sample of YSOs on the basis of the 0.8 - 12 $\mu$m spectral index ($\alpha$). Figure \ref{fig:AlphaHists} shows the proportion of high amplitude variables increasing with $\alpha$, noting the decrease in kurtosis for the later stage YSOs indicates that the distribution becomes more centrally dominant (and thus lower amplitude on average).

We can also examine the timescales of the observed variability, i.e., the durations of individual variable events. For our sample, we wish to identify any relationship between outburst duration and evolutionary stage, and thus we separate our sample of stars with $>1$ mag variability by the same spectral index groups as before. We use structure functions as an analogue to timescale, as laid out in \citet{2020SergisonSF} and \citet{2022LakelandSF}, although similar tests for YSO variability were carried out in \citet{20159FindeisenSF}.
\begin{equation}
    S(\tau) = \langle {[m(t) - m(t + \tau)]}^{2} \rangle
\end{equation}
Where $S$ is the structure, $\tau$ is the timescale between two points in a light curve, $m$ is the apparent magnitude, and $t$ is the modified Julian date of any given point in the light curve. This function produces a measure of correlation between each point in a light curve for a range of time baselines (we use 12 bins ranging from $\sim$100 days to $\sim$10 years). In Figure \ref{fig:AlphaStrucs}, we show that the signal is well correlated ($S$ is low) at the shortest timescales (here that is on the order of 3 months), and becomes less so as the timescale increases ($S$ is large). The implications here are that at shorter timescales it is easier to predict the next measurement in the time series (i.e. is less variable), and gets progressively more challenging at longer time baselines. It is worth noting that we do not see the higher $S$ values at the short timescales that \citet{2022LakelandSF} attributed to measurement error and short-term variability because our data from NEOWISE have been averaged at the shortest time baselines. We do see the same 'Knee' feature at large $\tau$ for 3 of the groups however, which those authors attributed to a combination of long-term linear trends or a maximum variability timescale for YSOs.

From both of the above tests it seems apparent that long-duration, high-amplitude variability is more common for more heavily embedded (likely younger) YSOs, as has been suggested previously \citep[see][]{Contreras2024}. 

\subsection{Outburst Colour Changes}\label{ssec:colours}

Most outbursting YSOs are observed to be bluer during an outburst, when compared against their pre-outburst colour. Within the optical regime, the blueing is most prominent in EXors \citep[see][]{Lorenzetti2012,2020SzEl_FUor}, for which the significantly increased accretion luminosity of the innermost part of the system dominates over the flux contribution from the outer disk (during outburst). This remains true (albeit to a lesser extent) for most FUors and other long-duration eruptive systems. The effect is less studied in the MIR, where the flux contribution from the cooler parts of the disk is greater.
\begin{figure}
    \includegraphics[width=0.9\columnwidth]{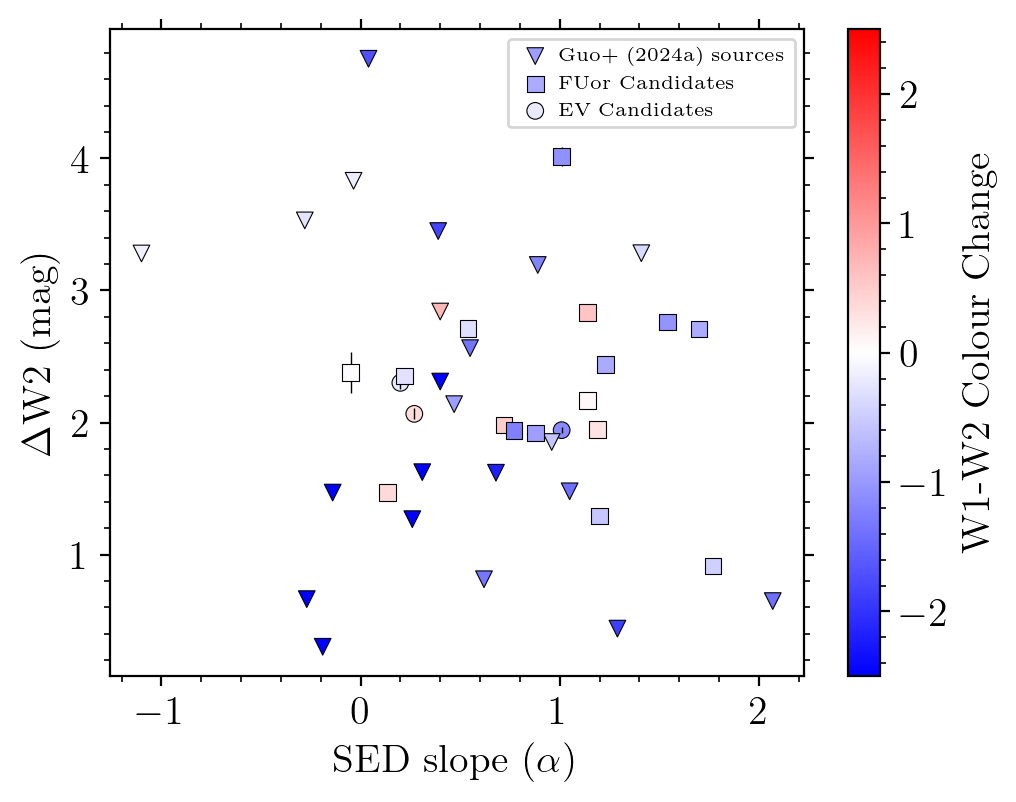}
    \includegraphics[width=0.9\columnwidth]{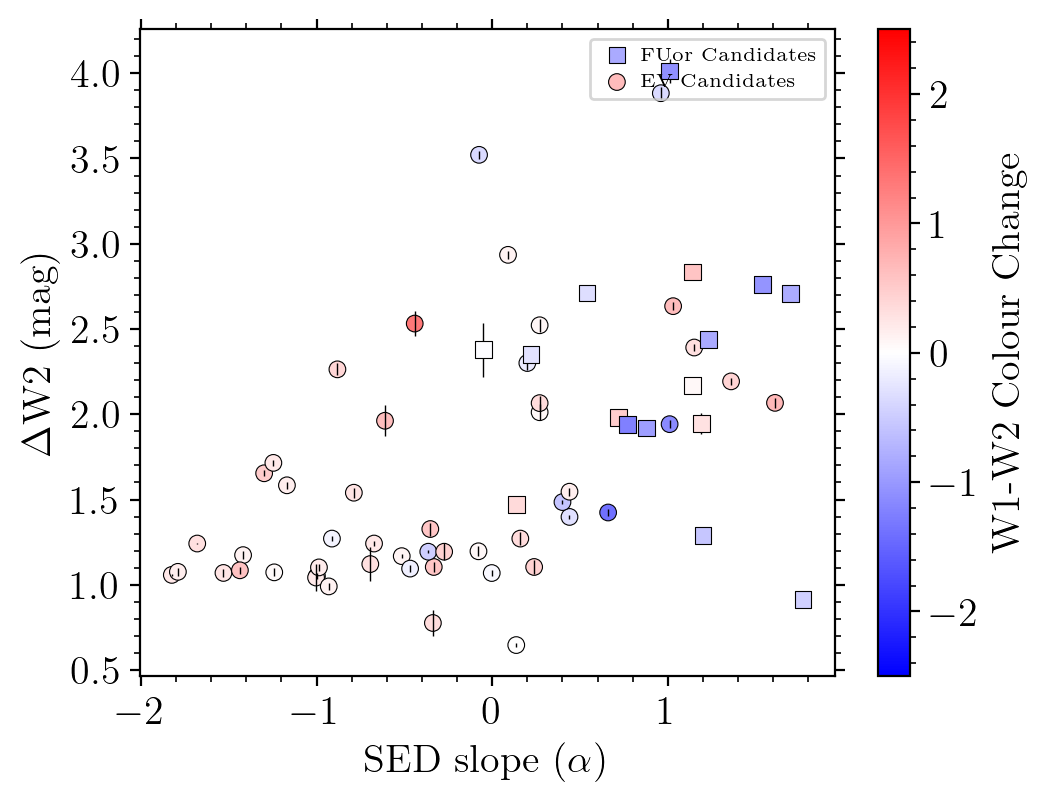}
    \caption{\textbf{ \textit{Top:} $W2$-bandpass amplitudes for the stars in the sample with outbursts labelled as IDEs or LDEs  (detected during outburst), against the pre-outburst SED slopes. The colour change during outburst is presented as the marker colour, with red points reddening during outburst and the opposite true for blue markers. White points represent stars with minimal colour change. Our sample of stars is represented by the circular and square points (for our uncertain and FUor candidates respectively). We include the NIR selected FUors (triangles) from \citet{2024Zhen_FU} to compare our YSOs with those selected from large amplitudes in the $K_s$ bandpass. The $\Delta W2$ values from that sample are based on lower limits for stars where there is no detection in $W2$.
    \textit{Bottom:} Similar plot for all the YSOs included in this work, regardless of outburst duration.}}
  \label{fig:AlphaColourAmps}
\end{figure}

We compared the $W2$ amplitudes with the SED class of the stars in our sample and a selection of long-duration eruptive YSOs, most of which were spectroscopically confirmed as FUors, from \citet{2024Zhen_FU}. Additionally, we investigated the colour change in $W1 - W2$, see Figure \ref{fig:AlphaColourAmps}. We have several broad findings:
\begin{itemize}
    \item Our full sample, see the lower panel of Figure \ref{fig:AlphaColourAmps}, has a typical behaviour of becoming redder when brighter. In particular, we note that YSOs with $[24]$ detections have larger changes in colour, having a median absolute colour change of 0.43 in $W1-W2$, compared to 0.23 for the sources from SPICY.
    \item We also note that the majority of our candidate FUors display the expected blueing during outbursts, in agreement with the long duration EV sample of \citet{2024Zhen_FU} (see the upper panel of Figure \ref{fig:AlphaColourAmps}). The colours of our mid-IR selected FUor candidates change less than the NIR-selected sources of those authors' sample however. 
    \item Our selection of Class II/flat-spectrum YSOs (from SPICY) are preferentially redder when brighter, which might be attributed to the fainter $W2$ magnitudes of these stars (in quiescence); this sample has a median quiescent brightness of $W1=12.46$ mag, compared to $11.31$ mag for the Class I sample. Given the logarithmic nature of the magnitude scale, a lower proportion of flux from enhanced accretion would be required to increase the magnitude of $W2$ than $W1$. The large YSO sample of \citet{Park_2021} (also using the NEOWISE time series) displayed a broad range of colour behaviours, similar to what we observe within our sample. Those authors attribute the behaviour to the interplay between accretion luminosity changes (bluer) as well as variable extinction from circumstellar material (redder), which can be moved during outburst.   
\end{itemize}

\section{Summary}

In this work, we have presented 68 potential candidate eruptive variable YSOs associated with the Cygnus X star-forming region, based on their decade-long mid-infrared NEOWISE light curves.

In total, we discovered 14 eruptive sources that we believe to be FUor candidates, which can be seen in Table \ref{tab:FUors}. Of these 14, eight have light curve morphologies similar to those of FUors or V1647 Ori-type stars. Three candidates show only a fading trend in their MIR light curves but have evidence for an initial rise in earlier NIR photometry. Two stars have extended brightening phases, comparable to V1515 Cyg. One final source has a currently ongoing outburst. In addition, six other sources are classified as plausible FUor candidates, with less clear light curves and more short-term variation.

We also locate a large number of YSOs with short-duration outbursts (some with EXor-like features) within both samples carried out, with 13 embedded Class I YSOs and 32 flat-spectrum or Class II stars. Compared with more evolved systems, we detected significantly more high amplitude (and eruptive) variables among the Class I YSOs (22.46\%).
This is especially true for the FUor type EVs, of which we have identified up to 16 candidate members, and several have a less common, slow-rising character. Given the similarities between these objects and the pre-outburst light curves of several novel FUors, these 'slow-risers' might be further examples of FUors that outburst at earlier times in redder wavelengths.

Given this potentially new population of eruptive young stars, follow-up observations are planned for those stars still at a state of high accretion rate and are thus bright in the near-infrared. This attempt may unveil any differences in accretion behaviour and disk/wind structure between the heavily embedded EVs and those sources selected via NIR/optical variability. 

Finally, we note the increased incidence of high amplitude variability for embedded YSOs (class I), as compared to less embedded sources (flat-spectrum and class II). We identified that the rate of variability of amplitudes that are $>$1 mag ranged from 2.9 times greater in $W1$, to 4.6 times greater in the $W2$ bandpass.

\section*{Acknowledgements}

ZG is supported by the ANID FONDECYT Postdoctoral program No. 3220029. This work was funded by ANID, Millennium Science Initiative, AIM23-0001. PWL acknowledges support from the UK's Science and Technology Facilities Council ST/Y000846/1. CM, PWL and NM recognise the computing infrastructure provided via STFC grant ST/R000905/1 at the University of Hertfordshire.
 
This publication makes use of data products from the Near-Earth Object Wide-field Infrared Survey Explorer (NEOWISE), which is a joint project of the Jet Propulsion Laboratory/California Institute of Technology and the University of Arizona. NEOWISE is funded by the National Aeronautics and Space Administration. This research has made use of the NASA/IPAC Infrared Science Archive, which is funded by the National Aeronautics and Space Administration and operated by the California Institute of Technology.

This work has made use of the University of Hertfordshire's high-performance computing facility (\url{http://uhhpc.herts.ac.uk}).
\section*{Data Availability}

The \textit{WISE} and \textit{Spitzer} data underlying this article are publicly available at the IRSA server \url{https://irsa.ipac.caltech.edu/Missions/wise.html}. ALLWISE/NEOWISE light curves are available at \url{http://star.herts.ac.uk/\textasciitilde cmorris/CygnusX\_YSOs/}. Reduced spectra are provided at \url{http://star.herts.ac.uk/~cmorris/}.



\bibliographystyle{mnras}
\bibliography{main_paper} 




\appendix
\section{Outburst Timescale fits with Emcee}\label{app:fits}

The plots below demonstrate some of the results of fitting a characteristic FUor-type shape to a NEOWISE light curve. As mentioned in the text (\S\ref{ssec:WISE_LCs}) we use \texttt{emcee} to fit 3 light curve components (representing the start of the outburst, the main rising slope, and the post-outburst linear decay) with 5 free parameters. Candidates were marked for further investigation if the reduced mean square error (RMSE) was less than 0.15. This value has been selected initially because it allows the code to recover all the stars in the sample that visual inspection identified as FUor candidates, with reasonably typical light-curves. The code does not recover the stars with consistent (and fairly slow) rising slopes, such as Source 1048. In a sample of stars of this size, our fitting technique is of questionable utility, however it will be of use when examining larger populations for which visual inspection would be inefficient. 

\begin{figure}
    \includegraphics[width=0.95\columnwidth]{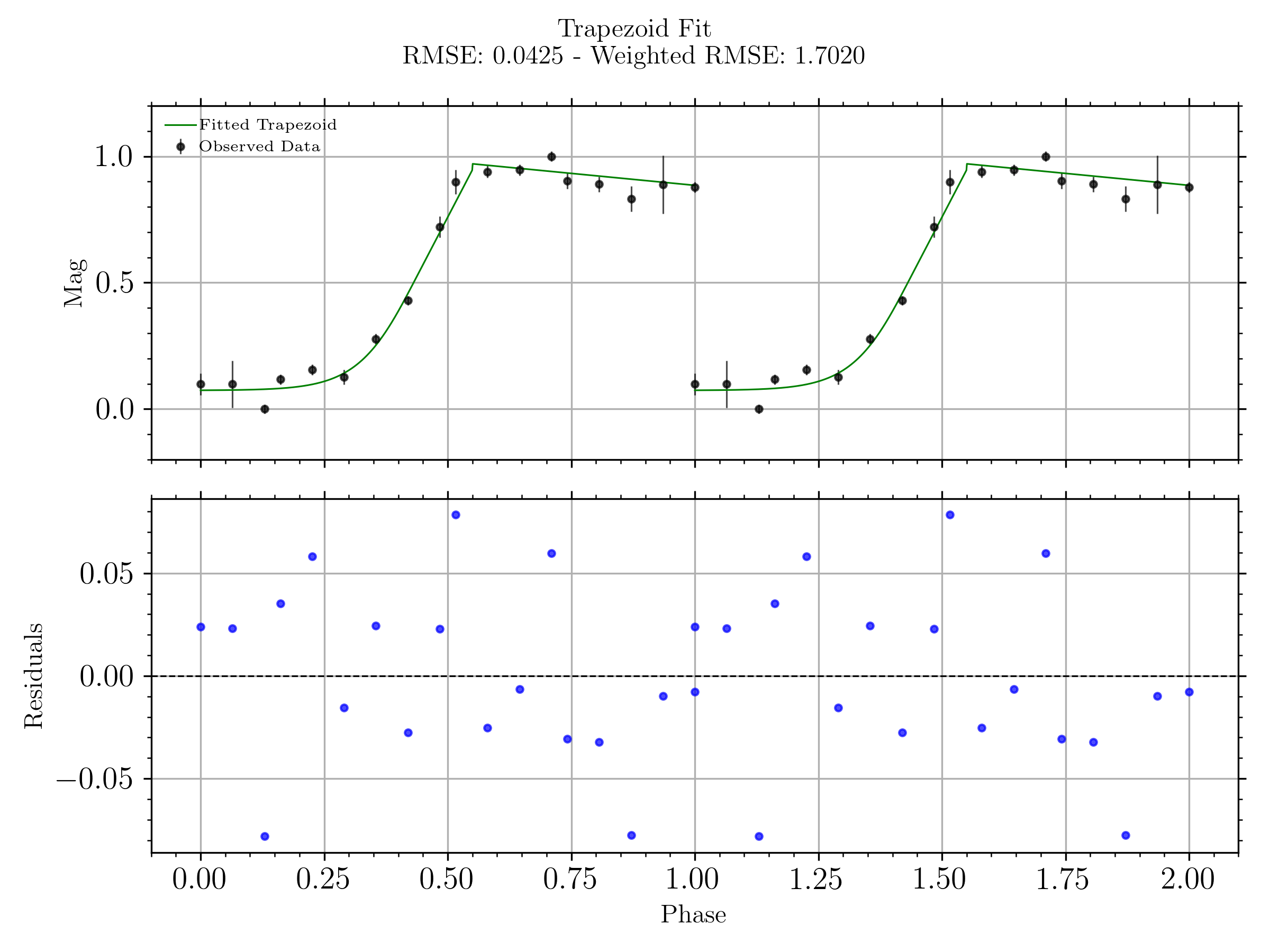}	
    \includegraphics[width=0.95\columnwidth]{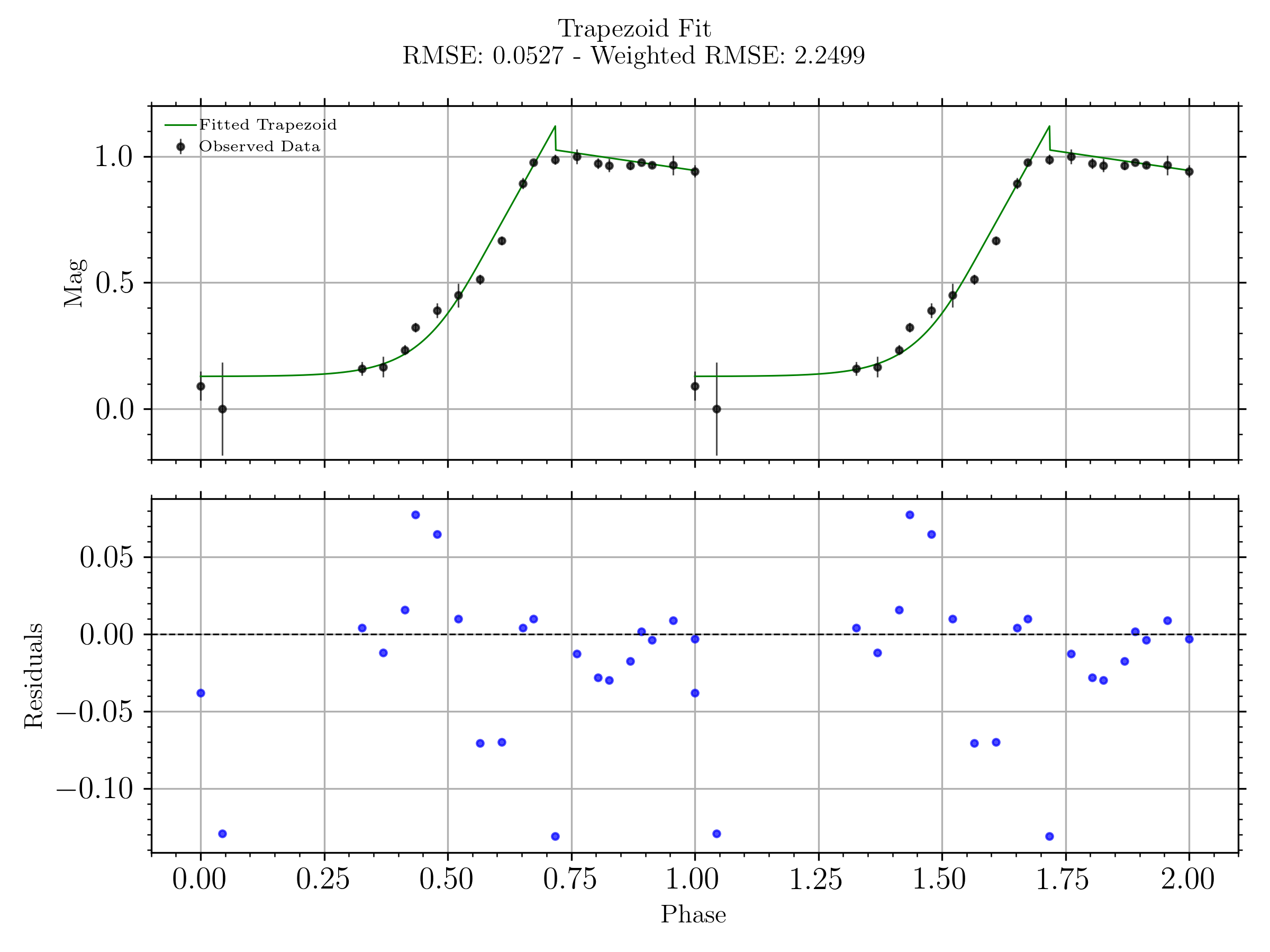}
    \caption{NEOWISE light curves (black points) and the best fit FUor outburst model (green line) for the first two FUor candidates mentioned in the text, Sources 257 (top) and 294 (bottom). The plots display the light curves twice, from zero to two phase, owing to the code's utility in additionally fitting to periodic signals.}
    \label{fig:YSOs_emcee_fit}
\end{figure}

\section{Additional Light Curves from the Embedded YSO Sample}\label{app:Kryu}
Listed in this section are the light curves from the stars not mentioned directly in the text. 
\begin{figure}
    \includegraphics[width=0.98\columnwidth]{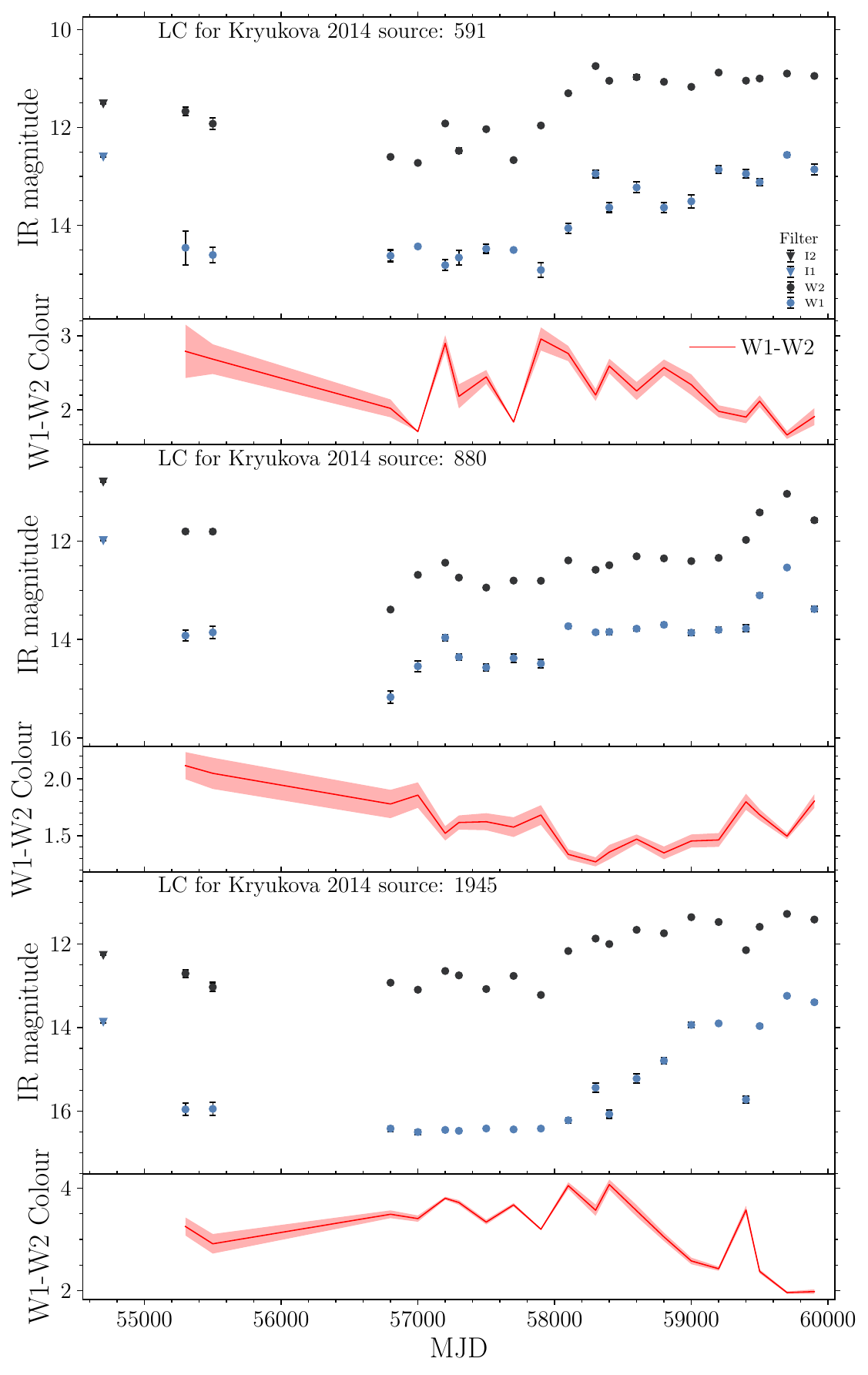}	
    \caption{NEOWISE $W1$ \& $W2$ light curves for the 3 long duration sources not detailed within the main text.}
    \label{fig:Kryu_LTEs_App}
\end{figure}
\begin{figure}
    \includegraphics[width=0.98\columnwidth]{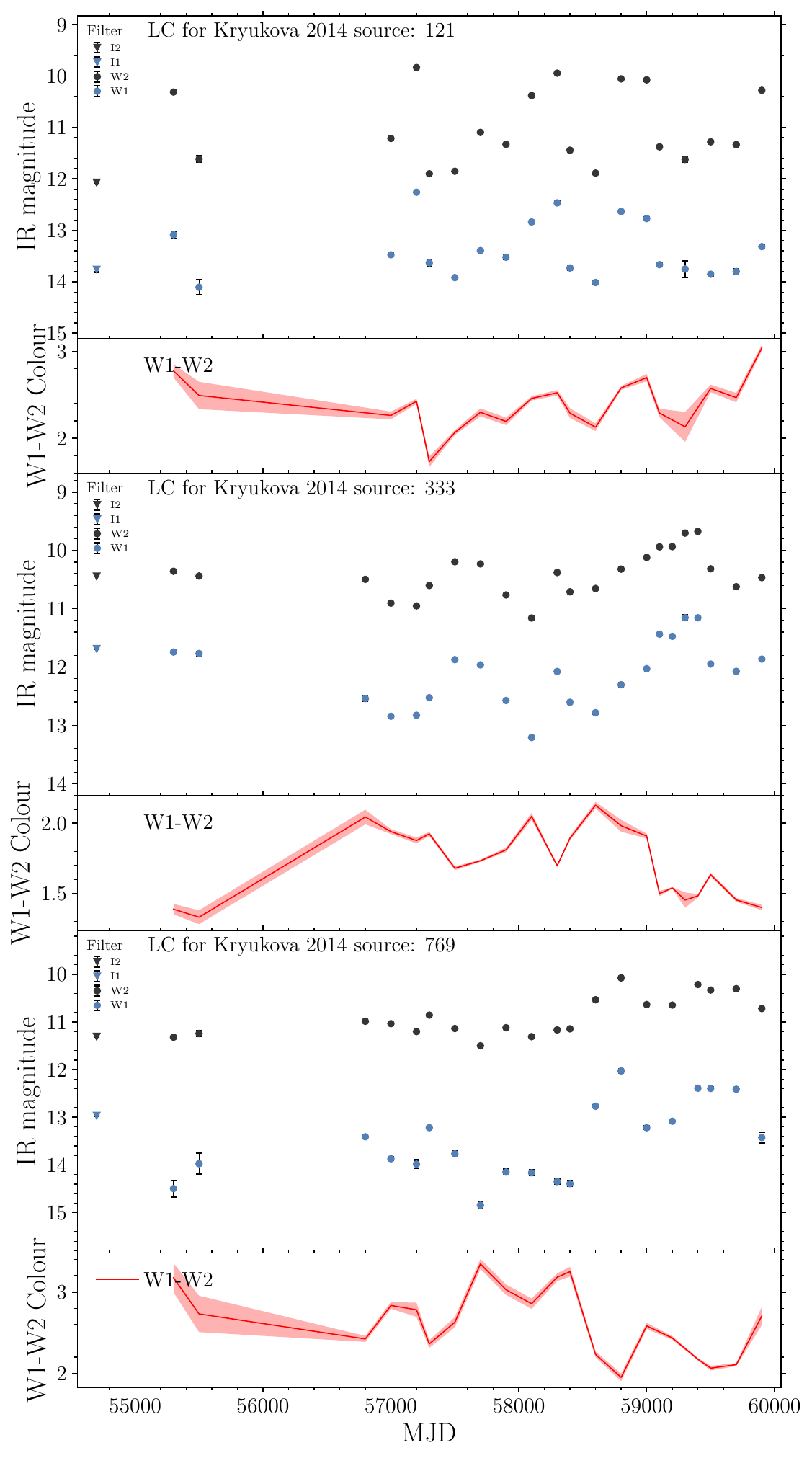}	
    \caption{NEOWISE $W1$ \& $W2$ light curves for the short-term variables not discussed in the main text.}
    \label{fig:Kryu_STEs_App}
\end{figure}

\section{Light Curves for high-amplitude Class I variable YSOs} \label{app:complex}
\begin{figure}
    \includegraphics[width=0.95\columnwidth]{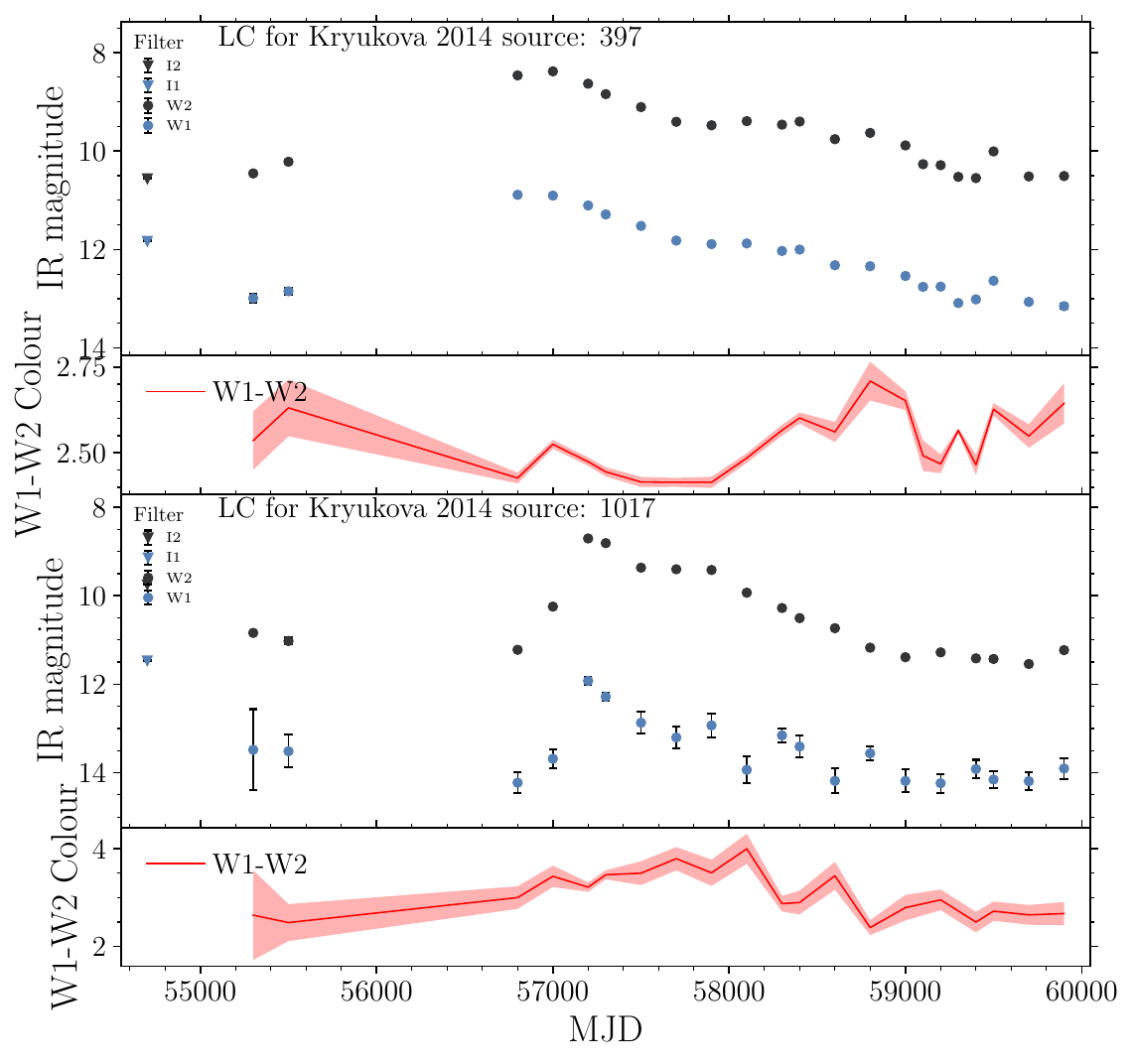}	
    \caption{NEOWISE $W1$ \& $W2$ light curves for two of the candidate eruptive variables with completed outbursts.}
    \label{fig:YSOs_kryu_IT}
\end{figure}

\begin{figure}
    \includegraphics[width=0.95\columnwidth]{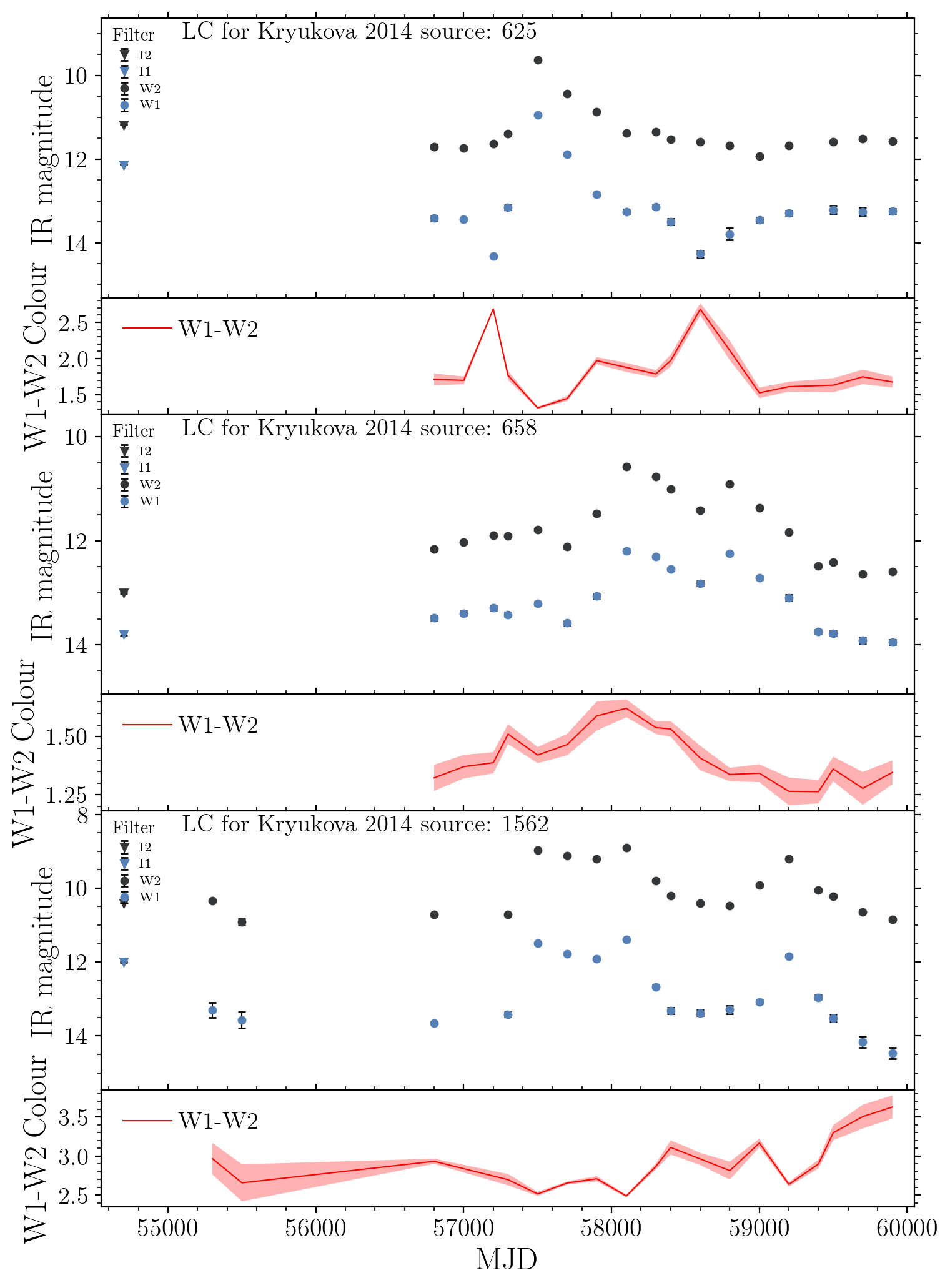}	
    \caption{NEOWISE $W1$ \& $W2$ light curves for three of the candidate eruptive variables with complex light curves.}
    \label{fig:YSOs_kryu_IT_busy}
\end{figure}
\begin{figure}
    \includegraphics[width=0.95\columnwidth]{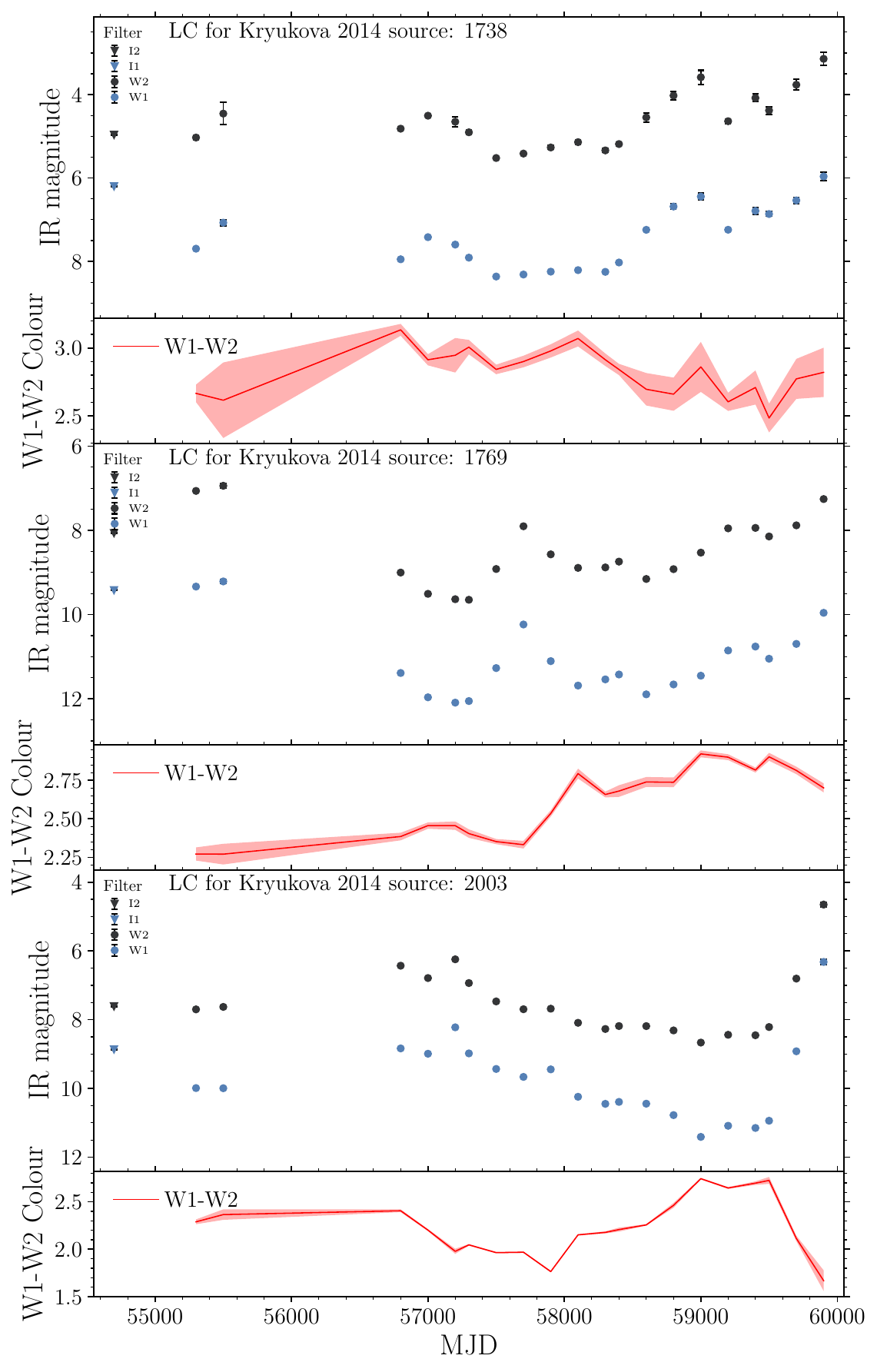}	
    \caption{NEOWISE $W1$ \& $W2$ light curves for the three stars with uncertain outburst durations.}
    \label{Fig:Ambig_YSOs}
\end{figure}

\section{Light Curves for High Amplitude variable YSOs from SPICY} \label{app:EXors}
Listed here are a selection of the NEOWISE MIR light curves for the EXor candidates discussed in section \ref{ssec:spicy_sources}. Light curves for all of the stars can be found in the online supplementary material.  

\begin{figure}
    \includegraphics[width=0.98\columnwidth]{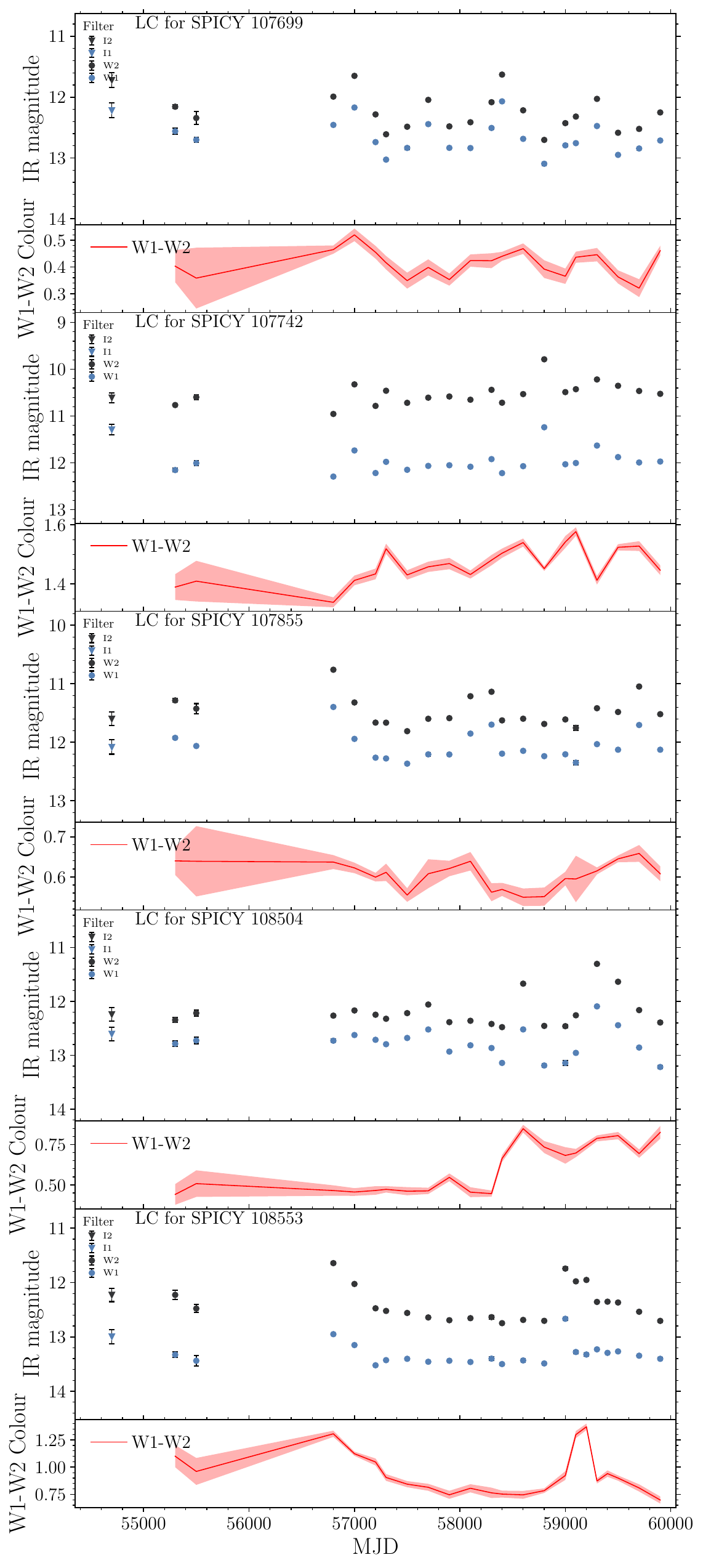}	
    \caption{NEOWISE $W1$ \& $W2$ light curves for the EXor candidates in the SPICY selected sample.}
    \label{fig:EXors1_spicy}
\end{figure}
\begin{figure}
    \includegraphics[width=0.98\columnwidth]{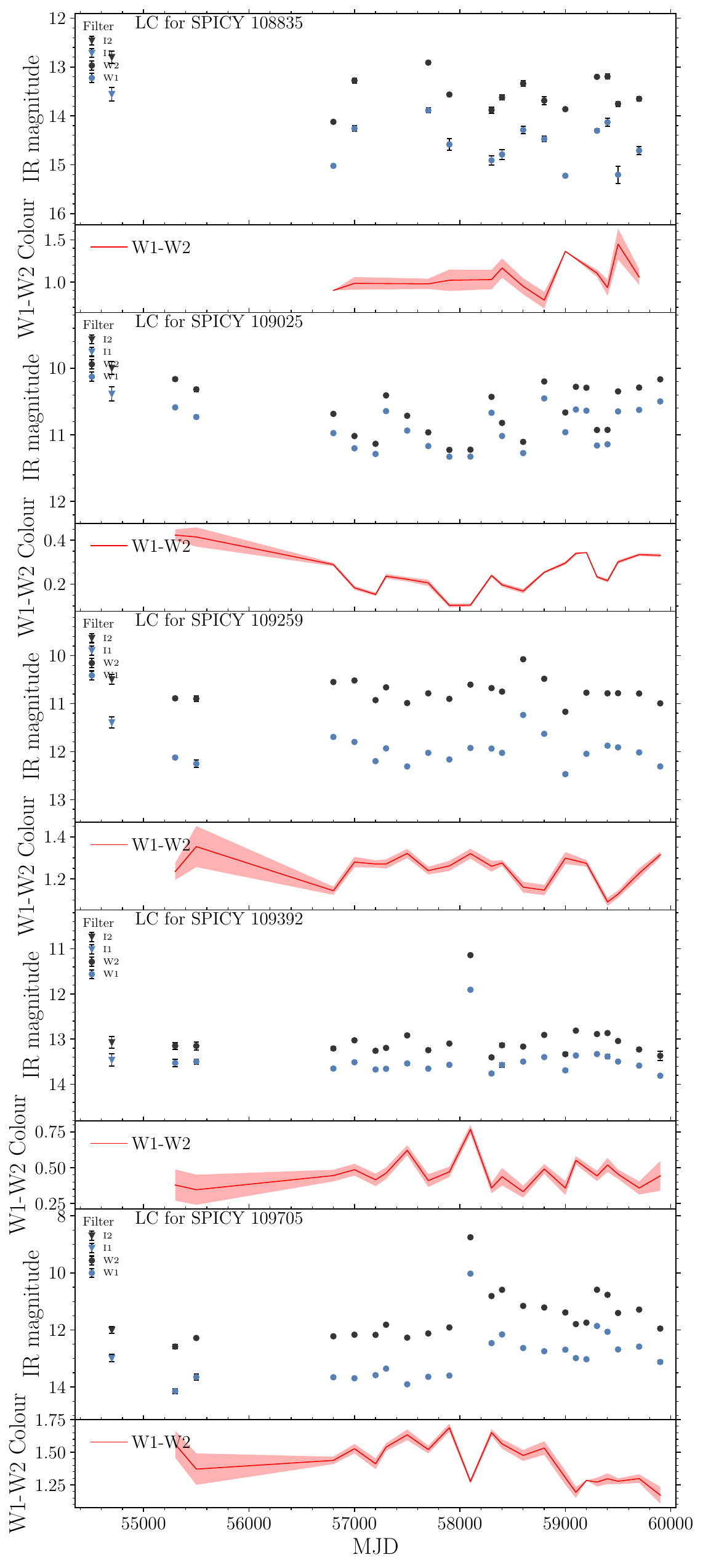}	
    \caption{NEOWISE $W1$ \& $W2$ light curves for the EXor candidates in the SPICY selected sample.}
    \label{fig:EXors2_spicy}
\end{figure}
\begin{figure}
    \includegraphics[width=0.98\columnwidth]{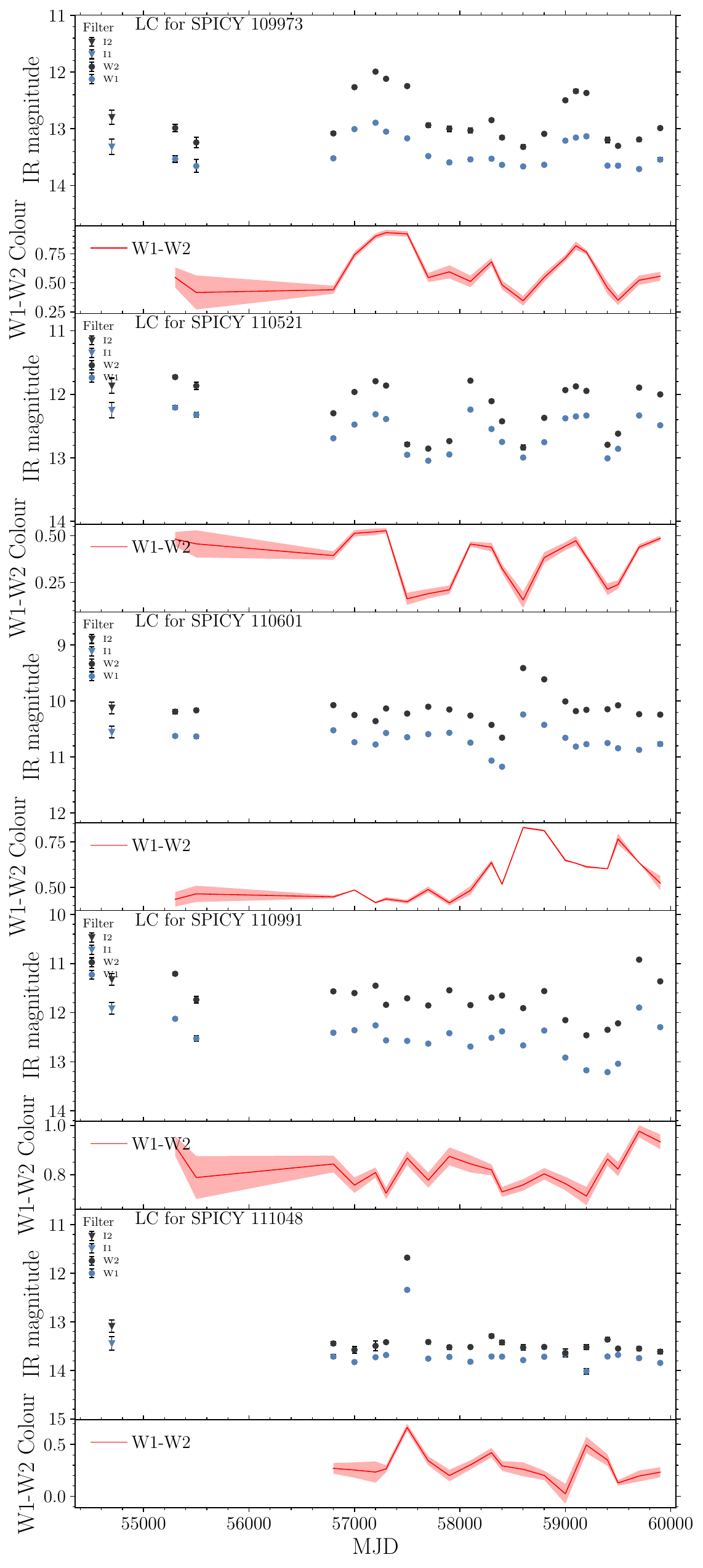}	
    \caption{NEOWISE $W1$ \& $W2$ light curves for the EXor candidates in the SPICY selected sample.}
    \label{fig:EXors3_spicy}
\end{figure}



\bsp	
\label{lastpage}
\end{document}